\documentclass[a4paper,12pt]{article}
\usepackage{amssymb,amsmath,amsfonts}

\usepackage[all]{xy}
\usepackage{pb-diagram,pb-xy}

\gdef\stackunder#1#2{\underset{#1}{#2}}

\def\endproof{\mbox{\ \rule{.1in}{.1in}}}

\begin{document}

\title{Supersymmetry and Homotopy}
\author{Serge MAUMARY and Izumi OJIMA \\
\\
{\small \textit{Institut de math\'ematiques, Universit\'e de Lausanne}}\\
{\small \textit{CH-1015 Lausanne, Switzerland}}\\
{\small \textit{and}}\\
{\small \textit{Research Institute for Mathematical Sciences}}\\
{\small \textit{Kyoto University, Kyoto 606-8502, Japan}}}
\date{}
\maketitle

\begin{abstract}
The homotopical information hidden in a supersymmetric structure is revealed
by considering deformations of a configuration manifold. This is in sharp
contrast to the usual standpoints such as Connes' programme where a
geometrical structure is rigidly fixed. For instance, we can relate
supersymmetries of types $N=2n$ and $N=(n,\ n)$ in spite of their gap due to
distinction between $\Bbb{Z}_2$(even-odd)- and integer-gradings.

Our approach goes beyond the theory of real homotopy due to Quillen,
Sullivan and Tanr\'e developed, respectively, in the 60's, 70's and 80's,
which exhibits real homotopy of a $1$-connected space out of its de
Rham-Fock complex with supersymmetry. Our main new step is based upon the
Taylor (super-)expansion and locality, which links differential geometry
with homotopy without the restriction of $1$-connectedness. While the
homotopy invariants treated so far in relation with supersymmetry are those
depending only on $\Bbb{Z}_2$-grading like the index, here we can detect new 
$\Bbb{N}$-graded homotopy invariants. While our setup adopted here is
(graded) commutative, it can be extended also to the non-commutative cases
in use of state germs (Haag-Ojima) corresponding to a Taylor expansion.
\end{abstract}

\section{The Algebraic Data of a Physical System}

The purpose of this section is to fix notation and our general viewpoint.
For simplicity, let us start with a classical system on a finite-dimensional
affine space $V$ with its coordinate space $V^{\ast }$. In reference to a
chosen probability measure $\omega $ on $V$, a polynomial algebra $\mathcal{F%
}^{0}=Sym(V^{\ast})$ on $V^{\ast }$ generates a sequence of function spaces, 
$\mathcal{F}^{0}\hookrightarrow \mathcal{F}_{0} =\mathcal{F}^{0}.\omega
^{1/2}\ \hookrightarrow L^{2}(V,\omega )=L^{2}(V,\omega )^{\ast
}\hookrightarrow (\mathcal{F}^{0})^{\ast }$, similar to the Gel'fand
triplet, where $\omega ^{1/2}$ denotes the half measure \cite{Nel70}
corresponding to $\omega $. The inner product of $\mathcal{F}_{0}$ is given
by $\langle f.\omega ^{1/2}\ |\ g.\omega ^{1/2}\rangle =\int_{V}\overline{%
f(x)}g(x)\omega (dx)$. By $\ \langle f,\ g.\omega ^{1/2}\rangle =\omega (%
\bar{f}g)$ a duality relation $(\mathcal{F}_{0})^{\ast }\stackrel{\omega }{%
\approx }\mathcal{F}^{0}$ holds. Hence we have $\mathcal{F}_{0}\stackrel{%
\omega }{\approx }Sym(V)$.

Then a self-dual algebra $P$ is defined by $P\equiv Sym(V\oplus V^{\ast })%
\stackrel{\omega }{\approx }\mathcal{F}_{0}\otimes \mathcal{F}^{0}$ which
has a canonical choice of a state determined by the Liouville measure $%
\omega $. Because of the above duality, a $^{\ast }$-involution can be
defined on $P$, through the action of which $\mathcal{F}_{0}$ and $\mathcal{F%
}^{0}$ are interchanged. A linear operator $A$ acting on $\mathcal{F}_{0},$ $%
A\in Hom(\mathcal{F}_{0},\mathcal{F}_{0})\approx \mathcal{F}_{0}\otimes 
\mathcal{F}^{0},$ and the corresponding one on $\mathcal{F}^{0},$ $A^{\ast
}\in Hom(\mathcal{F}^{0},\mathcal{F}^{0})\approx \mathcal{F}^{0}\otimes 
\mathcal{F}_{0},$ through $\omega $, have (Noether) charges, $%
Q_{A},Q_{A^{\ast }}\in P.$ Conversely, when the algebra $P$ is given, the
linear spaces $\mathcal{F}_{0},\mathcal{F}^{0}$ are reproduced from it as
induced representations from the trivial one on $V^{\ast }$ or on $V$ (via $%
\omega $).

According to the standard procedure, the classical dynamics on $\mathcal{F}%
_{0},\mathcal{F}^{0}$ is described with the Poisson-Lie bracket on $P$ which
is closely related with the quantization $\frak{U(}\hslash eis(V))$ through $%
P[\hslash ]\approx \frak{U(}\hslash eis(V))$\-\-. Here $\hslash eis(V)\equiv
V\oplus \hslash \Bbb{C}\oplus V^{\ast }$ denotes the Heisenberg-Lie algebra
equipped with the Lie bracket coming from the evaluation map $V\times
V^{\ast }\rightarrow \Bbb{C}$. Then a dynamics $\partial _{t}$ on $V$ has a
Hamiltonian $H\in P$ as its Noether charge such that $H\}$ defined on $P$ by 
$(H\})f\equiv \{f,H\}$ implements $\partial _{t}:\partial _{t}f=\{f,H\}$.
Similarly any symmetry of the dynamics $\partial _{t}$ constituting a
finite-dimensional Lie algebra $\frak{g}$ has a Noether charge in the
Poisson-Yang algebra $V\oplus \frak{g}^{\ast }\oplus V^{\ast }$ which is a
generalization of Heisenberg-Lie algebra.

In addition to the bosonic variables belonging to $\mathcal{F}^{0}$, there
are fermionic ones constituting a $\mathcal{F}^{0}$-module $L^{0}$, which
are nilpotent $\psi ^{2}=0$ for $\psi \in L^{0}$. Then the above structure
survives in an extended form with the evaluation superbracket on $%
L_{0}\oplus \hslash eis(V)\oplus L^{0}=L_{0}\oplus (V\oplus \hslash \Bbb{C}%
\oplus V^{\ast })\oplus L^{0}$ where $L^{0}$ and $L_{0}=(L^{0})^{\ast }$ are 
$\hslash eis(V)$-modules graded by $+1$ and $-1$, respectively. Again, $%
\mathcal{F=F}^{0}\mathcal{\otimes F}^{\geq 1}$ is induced from the
Poisson-Lie superalgebra $P\otimes (\wedge L^{0})\otimes (\wedge L_{0})$,
where $\mathcal{F}^{0}=Sym(V^{\ast }),\ \mathcal{F}^{1}=\mathcal{F}%
^{0}\otimes L^{0}$ and a linear operator $\mathcal{F}^{k}\mathcal{%
\rightarrow F}^{k+1}$ is implemented by a degree $1$ Noether charge $Q\in
P\otimes (\wedge L_{0})\otimes (\wedge L^{0}).$

This simple physical picture can be extended to a general manifold by
considering the de Rham complex $\Omega $ on it, but not for charges. In
fact, its basic algebra $\mathcal{F}^{0}=\Omega ^{0}$ consists of
differentiable functions, with their commutative pointwise product. The
symmetries are given by the (infinite-dimensional) Lie algebra $\mathcal{X}%
\simeq Der(\Omega ^{0})$ and differentiable $1$-forms $\varphi =\sum
f_{i}dg_{i}:\mathcal{X}\ \backepsilon \ X\ \longmapsto \sum f_{i}[X,\ g_{i}]$
constitute the space ${\Omega }^{1}$ of fermionic generators. It is a
finitely generated projective $\Omega ^{0}$-module (equivalent to a finite
direct sum of $\Omega ^{0}$), $\Omega ^{0}$-dual to the space $\Omega _{-1}$
of the differentiable $1$-currents. Namely, a choice of an orientation ${%
\omega }_{-n}$ and a differentiable oriented volume ${\Phi }^{n}$ on the
orientable double covering of our manifold with dimensionality $n$
determines a differentiable $0$-current ${\omega }={\Phi }^{n}.{\omega }%
_{-n} $ dual to ${\Omega }^{0}$, and also differentiable ($-1$)-currents $%
c_{X}$ dual to ${\Omega }^{1}$ : $\langle c_{X},\ \varphi \rangle _{\omega
}=\omega ({\iota }_{X}(\varphi ))=\langle {\omega }_{-n},\ {\Phi }^{n}\wedge 
{\iota }_{X}(\varphi )\rangle $ where ${\iota }_{X}$ is the evaluation. Then
the space ${\Omega }_{-1}$ of differentiable ($-1$)-currents is identified
with ${\Omega }^{n-1}.{\omega }_{-n}$. Because of ${\Omega }^{n}={\Omega }%
^{0}{\Phi }^{n}$ and of ${\Omega }^{n-1}=\mathrm{Lin}\{{\iota }_{X}(\Phi
^{n});\ X\in \mathcal{X}\}$, ${\Omega }^{1}$ is both ${\Omega }^{0}$-dual
and $\Bbb{R}$-dual to ${\Omega }_{-1}$, respectively, by $\langle c_{X},\
\varphi \rangle ={\iota }_{X}(\varphi )$ and by $\langle c_{X},\ \varphi
\rangle _{\omega }=\omega ({\iota }_{X}(\varphi ))$, while ${\Omega }_{-1}$
is a regular representation of the Lie algebra $\mathcal{X}$. Then we have
the ${\Omega }^{0}$-Heisenberg algebra ${\Omega }_{-1}\oplus \hslash {\Omega 
}^{0}\oplus {\Omega }^{1}$, with ${\Omega }^{0}$-bilinear evaluation
superbracket. It is a deformation of the Poisson superalgebra $P=(\wedge _{{%
\Omega }^{0}}{\Omega }_{-1})\otimes _{{\Omega }^{0}}(\wedge _{{\Omega }^{0}}{%
\Omega }^{1})$. The latter induces the de Rham complexes ${\Omega }_{-},{%
\Omega }^{+}$ of differentiable currents and forms.

Taking $X_{i}$ as fixed and a $p$-form $\varphi $ as variable, ${\Omega }%
^{p}\backepsilon \ \varphi \longmapsto \langle \varphi (X_{1},\cdots
,X_{p})\rangle _{\omega }=\omega ((\iota _{X_{p}}\circ \cdots \circ \iota
_{X_{1}})(\varphi ))\equiv \langle c_{X_{1}\wedge \cdots \wedge X_{p}},\
\varphi \rangle _{\omega }$, we obtain a differentiable exterior $(-p)$%
-current $c_{X_{1}\wedge \cdots \wedge X_{p}}\in \Omega _{-p}$ similarly to
the above case of $p=1$. When we choose a suitable sequence of ${\omega }%
^{(k)}$ of differentiable $0$-currents tending to a Dirac measure $\delta
_{z}$ at $z\in Z$, ${\omega }^{(k)}\stackunder{k\rightarrow \infty }{%
\rightarrow }\delta _{z}$, the limiting formula is given by $\langle \varphi
(X_{1},\cdots ,X_{p})\rangle _{\omega }=\delta _{z}(\varphi (X_{1},\cdots
,X_{p}))
=\stackunder{_{i_{1}\cdots i_{p}}}{\sum }\varphi _{i_{1}\cdots i_{p}}(z)
X_{1}^{i_{1}}(z)\cdots X_{p}^{i_{p}}(z)$. Comparing this with $\langle
c_{-p},\varphi \rangle \equiv \int{}_{c_{-p}}\varphi =\int_{c_{-p}}%
\stackunder{_{i_{1}\cdots i_{p}}}{\sum }\varphi _{_{i_{1}\cdots
i_{p}}}dz^{i_{1}}\cdots dz^{i_{p}}$, we can regard $X_{1}(z)\wedge \cdots
\wedge X_{p}(z)$ as an infinitesimal version\textit{\ }of (an integral over)
a cycle $c_{-p}$ at a point $z\in Z$, which generalizes a Dirac measure $%
\delta _{z}$ to a current $c_{X_{1}\wedge \cdots \wedge X_{p}}\in \Omega
_{-p}$. The point-like nature of a current $c_{X_{1}\wedge \cdots \wedge
X_{p}}\in \Omega _{-p}$ and the $\Omega ^{0}$-linearity of the form $\varphi 
$ and of $\stackunder{\Omega ^{0}}{\wedge }$ can be recovered in this form, $%
\varphi (fX_{1}\stackunder{\Omega ^{0}}{\wedge }\cdots \stackunder{\Omega
^{0}}{\wedge }X_{p})=\varphi (X_{1}\stackunder{\Omega ^{0}}{\wedge }fX_{2}%
\stackunder{\Omega ^{0}}{\wedge }\cdots \stackunder{\Omega ^{0}}{\wedge }%
X_{p})=\cdots =f\varphi (X_{1}\stackunder{\Omega ^{0}}{\wedge }\cdots 
\stackunder{\Omega ^{0}}{\wedge }X_{p})$ for $f\in \Omega ^{0}$ : $c(\varphi
)=\int \langle c_{z},\varphi _{z}\rangle \omega $ with $z\longmapsto \langle
c_{z},\varphi _{z}\rangle \in \Omega ^{0}$.

As the symmetries $\in \mathcal{X}$ are not $\Omega ^{0}$-linear but $\Omega
^{0}$-Leibniz, they have no Noether charge in $P\otimes \mathcal{X}^{\ast }$
(Poisson-Yang). In the case of invariant de Rham complex on a Lie group $G$
with Lie algebra $\frak{g}$, the Heisenberg algebra becomes $\frak{g}%
_{-1}\oplus \hslash \Bbb{C}\oplus \frak{g}^{1}$ and $d:\wedge ^{k}\frak{%
g\rightarrow }\wedge ^{k+1}\frak{g}$ has a Noether charge. In general, this
problem will be solved by Taylor superexpansions: They provide a Heisenberg
algebra $L_{-}\oplus \hslash \Bbb{C}\oplus L^{+}$, where $L_{-}$ is
associated with a Lie-algebraic cochain complex $L=\{L_{-k}\}_{k\in \Bbb{N}%
_{0}}$ ($\Bbb{N}_{0}\equiv \Bbb{N}\cup \{0\}$) (see Definition 3 in Sect.2)
augmented by $L_{0}\rightarrow V$: 
\begin{equation}
L(V):\ \cdots \rightarrow L_{-2}\rightarrow L_{-1}\rightarrow
L_{0}\rightarrow V\rightarrow 0,
\end{equation}
with degree shift by $-1$, and $L^{+}=L_{-}^{\ast }$. $L_{-}$ provides a
regular representation of $L(V)$. Then a symmetry given by a Lie algebra $%
\frak{g}$ will have a Noether charge in $P\otimes \frak{g}^{\ast }$.

In the general setting with the implementer $H\in P$ of a derivation $\delta 
$ generating a dynamics, an ordinary symmetry is described by an element $%
X\in \frak{g}^{\ast }$ whose charge $Q_{X}$ (of degree $0$) satisfies $%
[Q_{X},H]=0$. A supersymmetry is given by an operator of degree $1$ acting
on $\mathcal{F}^{+}$ or $\mathcal{F}_{-}$ defined by $\mathcal{F}^{+}\equiv 
\mathcal{A}_{c}(\overline{L^{+}})$ and $\mathcal{F}_{-}\equiv (\mathcal{F}%
^{+})^{\ast }$ (see Sect.2) whose charge $Q$ (of degree $1$) also satisfies $%
[Q,H]=0$. These (super)symmetries form a canonical Lie superalgebra, either
with $\Bbb{Z}$ (or $\Bbb{N}$)-grading or $\Bbb{Z}_{2}$-grading. One of our
main objective here is to examine the possibility of passing from $\Bbb{Z}%
_{2}$-grading to $\Bbb{Z}$ (or $\Bbb{N}$)-grading, by asking when
(super)symmetries can be realized by nilpotent charges of degree $1$.

This paper is organized as follows: The essence of algebraic formulation of
homotopy is explained in Sect.2, which is extended in Sect.3 to the
equivariant situation involving a principal bundle. Then, the mutual
relation between de Rham complexes and homotopy is explained in Sect.4 by
using the Taylor superexpansions, a graded extension of Taylor expansions.
In Sect.5, the relation between the complexes for the total space and for
the base space of the bundle is clarified. The method of algebraic homotopy
is generalized in Sect.6 to the non 1-connected cases. By using the
developed techniques, the obstruction to transform $\Bbb{Z}_2$-grading to $%
\Bbb{Z}$ (or $\Bbb{N}$)-grading caused by the presence of torsion is shown
to be eliminated by a homotopy covering.

\section{Algebraic Homotopy}

To facilitate the unified treatment of chain and cochain complexes, we note
the following simple observation: A chain complex $C=\{C_{k}\}_{k\in \Bbb{N}%
_{0}}$ over a ground ring $K$, 
\begin{equation}
C:\cdots \stackrel{\partial }{\longrightarrow }C_{k}\stackrel{\partial }{%
\longrightarrow }C_{k-1}\stackrel{\partial }{\longrightarrow }\cdots 
\stackrel{\partial }{\longrightarrow }C_{0}\stackrel{\partial }{%
\longrightarrow }0,
\end{equation}
consisting of graded $K$-modules $C_{k}$ with $K$-linear boundaries $%
\partial =\partial _{k}:$ $C_{k}\rightarrow C_{k-1}$ (of degree $-1$ and
satisfying $\partial ^{2}=0$) can be viewed as a \textit{cochain} complex $%
\mathcal{F}_{-}=\{\mathcal{F}_{-k}\}$ of \textit{negatively graded} $K$%
-modules, simply by a renaming $\mathcal{F}_{-k}\equiv C_{k}$ to reverse the
sign of degrees which makes $\partial :$ $\mathcal{F}_{-k}\rightarrow 
\mathcal{F}_{-k+1}$ to be coboundaries of degree $1$: 
\begin{equation}
\mathcal{F}_{-}:\cdots \stackrel{\partial }{\longrightarrow }\mathcal{F}%
_{-k} \stackrel{\partial }{\longrightarrow }\mathcal{F}_{-k+1}\stackrel{%
\partial }{\longrightarrow }\cdots \stackrel{\partial }{\longrightarrow }%
\mathcal{F}_{0} \stackrel{\partial }{\longrightarrow }0.
\end{equation}
In spite of its degree $+1$, we keep the name \textit{boundary} for $%
\partial :$ $\mathcal{F}_{-k}\rightarrow \mathcal{F}_{-k+1}$, because it
appears only in \textit{negatively graded} complexes without any confusions.

Dual to this is a \textit{positively graded} cochain complex, $\mathcal{F}%
^{+}=\{\mathcal{F}^{k}\}$, defined by $\mathcal{F}^{k}\equiv (\mathcal{F}%
_{-k})^{\ast }$ with $K$-linear coboundaries $d=(-)^{\bullet }{\partial }%
^{\ast }$ of degree $1$: 
\begin{equation}
\mathcal{F}^{+}:0\stackrel{d}{\longrightarrow }\mathcal{F}^{0}\stackrel{d}{%
\longrightarrow }\ \cdots \stackrel{d}{\longrightarrow }\mathcal{F}^{k-1}%
\stackrel{d}{\longrightarrow }\mathcal{F}^{k}\stackrel{d}{\longrightarrow }%
\cdots .
\end{equation}
To formulate the duality between $\mathcal{F}^{+}=(\mathcal{F}_{-})^{\ast }$
and $\mathcal{F}_{-}$ in a consistent and convenient way, this simple trick
plays a crucial role, by which both $\mathcal{F}^{+}$ and $\mathcal{F}_{-}$
can be treated as cochain complexes on the same ground. Without this it is
difficult, for instance, to construct a (universal Poisson) cochain complex $%
\mathcal{F}_{-}^{\prime }\otimes \mathcal{F}^{+}$ for cochain maps $\in Hom(%
\mathcal{F}_{-},\mathcal{F}_{-}^{\prime })$ between cochains $\mathcal{F}%
_{-}\rightarrow \mathcal{F}_{-}^{\prime }$. Here the tensor product $%
\mathcal{F}_{-}^{\prime }\otimes \mathcal{F}^{+}$ of two cochain complexes
is a cochain complex equipped with the coboundary $D=\partial ^{\prime
}\otimes I+(-1)^{\bullet }I\otimes d$ and the total $\Bbb{Z}$-grading
defined by $(\mathcal{F}_{-}^{\prime }\otimes \mathcal{F}^{+})_{k}=%
\stackunder{l+m=k,\ l\leq 0,\ m\geq 0}{\oplus }\mathcal{F}_{l}^{\prime
}\otimes \mathcal{F}^{m}$. \bigskip A cochain map $\phi $ of degree $k$ is
defined by a family $\{\phi _{-l}\}$ of $K$-linear maps $\phi _{-l}:\mathcal{%
F}_{-l}\rightarrow \mathcal{F}_{-l+k}^{\prime }$ satisfying the condition 
\begin{equation}
\lbrack Q,\phi ]\equiv \partial ^{\prime }\circ \phi -(-)^{k}\phi \circ
\partial =0,  \label{Der}
\end{equation}
where $Q$ denotes a universal charge for coboundary. By definition, the
element $\tilde{\phi}$ in $P=\mathcal{F}_{-}^{\prime }\otimes \mathcal{F}%
^{+} $ corresponding to $\phi $ has non-vanishing components only at degree $%
k$ in $\stackunder{l+m=k,\ l\leq 0,\ m\geq 0}{\oplus }\mathcal{F}%
_{l}^{\prime }\otimes \mathcal{F}^{m}=\stackunder{l\leq 0}{\oplus }Hom(%
\mathcal{F}_{-l},\mathcal{F}_{-l+k}^{\prime })$. When $\phi $ has an
algebraic implementer $\tilde{\phi}=\sum c^{\prime }\otimes f$ (Noether
charge) in $P=\mathcal{F}_{-}^{\prime }\otimes \mathcal{F}^{+}$, the
condition (\ref{Der}) on $\phi $ amounts to $D(\tilde{\phi})=0$. Actually,
the map $\mathcal{F}_{-}^{\prime }\otimes \mathcal{F}^{+}\rightarrow Hom(%
\mathcal{F}_{-},\mathcal{F}_{-}^{\prime })$ sending $c^{\prime }\otimes f$
to $c^{\prime }f(\cdot )\ $ is a $0$-degree cochain map. The above equation $%
D(\tilde{\phi})=0$ can be interpreted as the invariance of the charge $%
\tilde{\phi}$: When $\phi $ describes an ordinary infinitesimal symmetry (of
degree 0) $\mathcal{F}_{-l}\rightarrow \mathcal{F}_{-l}$ on a self dual
cochain, the condition $D(\tilde{\phi})=0$ is seen to be equivalent to each
of $[\phi ,\partial ]=0$ and $[\phi ,\partial ^{\ast }]=0$ and hence, to $%
[\phi ,H]=0$ where $H=\partial \partial ^{\ast }+\partial ^{\ast }\partial $.

To emphasize the importance of integer grading (in contrast to $\Bbb{Z}_2$%
-grading) we mention the following examples of chain complexes:

1) The chain complex $C(\mathcal{C})$ of a category $\mathcal{C}$, based in
each degree $k\geq 0$ on the sequences of $k$ arrows $\rightarrow
\rightarrow \rightarrow \cdots $, with $\partial =\sum (-)^{i}\partial _{i},
\ \partial _{i}$ expressing the composition of morphisms, and $\partial
^{2}=0$ their associativity. The $\partial _{i}$ lead to an intermediate
topological space $B\mathcal{C}$, the so-called classifying one.

2) The infinitesimal version of the above (with half open arrows) is
expressed by de Rham differentiable currents.

3) The corresponding equivariant cases are expressed by the Cartan-de Rham
currents $\wedge \frak{g}_{-1}$, where $\frak{g}_{-1}$ is the regular
representation space of a Lie algebra $\frak{g}$ with degree -1 assigned.
This corresponds to the equivalence between representation space $V$ of ${%
\frak{g}}$ and Lie structure on $V\oplus \frak{g}$ with $0$ bracket on $V$.
There are two different boundaries $\partial $ extending the bracket $[\cdot
\ ,\ \cdot ]:\frak{g}_{-1}\wedge \frak{g}_{-1}\rightarrow \frak{g}_{-1}$,
one as derivation for the algebra structure and the other as derivation for
the coalgebra structure.

We proceed now to introduce in a new way the notion of algebraic homotopy
based upon the general notion of freeness in the categorical formulation. We
recall that, for a given functor $F^{\prime }:\mathcal{C}\rightarrow 
\mathcal{C}^{\prime }$ from a category $\mathcal{C}$ to another one $%
\mathcal{C}^{\prime }$, an object $X$ in $\mathcal{C}$ is called $\mathcal{C}%
^{\prime }$-\textit{free} if one can find an object $X^{\prime }$ in $%
\mathcal{C^{\prime }}$ and a natural isomorphim 
\begin{equation}
\mathcal{C}(X,Y)\approx \mathcal{C}^{\prime }(X^{\prime },F^{\prime }Y)\quad %
\mbox{\rm for}\ \forall Y,
\end{equation}
Such an $X^{\prime }$ is called a \textit{basis} for $X$, and each morphism
in $\mathcal{C}$ [called ``$\mathcal{C}$-map'' hereafter for short] from $X$
corresponds exactly to a $\mathcal{C}^{\prime }$-map from the basis $%
X^{\prime }$. In standard situations where $\mathcal{C}$ consists in
monoidal objects in a category $\mathcal{C}^{\prime \prime }$ underlying $%
\mathcal{C}^{\prime }$, any object $X^{\prime }$ serves as basis for a free
object $X$ constructed formally. In such cases, we have a functor $F:%
\mathcal{C}^{\prime }\rightarrow \mathcal{C}$ called a (left) adjoint to $%
F^{\prime }$, 
\begin{equation}
\mathcal{C}\ \stackrel{F^{\prime }}{\stackunder{F}{\rightleftarrows }}\ \ 
\mathcal{C}^{\prime },
\end{equation}
creating a free object $X=F(X^{\prime })$ in $\mathcal{C}$ from each basis $%
X^{\prime }$ in $\mathcal{C}^{\prime }$ and a natural isomorphism 
\begin{equation}
\mathcal{C}(FX^{\prime },Y)\approx \mathcal{C}^{\prime }(X^{\prime
},F^{\prime }Y)\quad \mbox{\rm for}\ \forall Y.
\end{equation}
Then, the canonical maps, $X^{\prime }\stackrel{\eta _{X^{\prime }}}{%
\rightarrow }F^{\prime }FX^{\prime },\ FF^{\prime }X\stackrel{\epsilon _{X}}{%
\rightarrow }X$ (called unit and counit, respectively), generate resolutions
such that $\epsilon _{FX^{\prime }}F(\eta _{X^{\prime }})=I_{FX^{\prime }},\
F^{\prime }(\epsilon _{X})\eta _{F^{\prime }X}=I_{F^{\prime }X}$. So any
object $X\in \mathcal{C}$ has a free presentation $FF^{\prime }X\stackrel{%
\epsilon _{X}}{\rightarrow }X\approx $ $FF^{\prime }X\diagup RX$ with an
equivalence relation $RX$ (which is, in an abelian category, nothing but the
equivalence w.r.t. $Ker(\epsilon _{X})$). Dually, the adjunction defines
co-free objects $F^{\prime }Y$. Adjunction of a category $\mathcal{C}$ with
a point category, consisting of one object with its identity, plays an
important role. A corresponding free (resp., cofree) object is initial
(resp., final) in $\mathcal{C}$. The arrows linking them to other objects
correspond bijectively to these objects. In a simplex category, all objects
are initial (resp., final). Groups give examples of such a simplex category,
with elements of the group as objects, and arrows $\mu :\gamma \longmapsto
\mu \gamma $. Also Hilbert spaces give such example, with objects the lines
generated by unit vectors $X$ and morphism $X{\rightarrow }Y$ the orthogonal
projection: $\mathcal{C}(X,Y)\approx \langle X,Y\rangle $. By adding the $0$
object, partial isometries provide adjoint functors. The existence of an
adjoint functor ($F$ in the above) to a given one ($F^{\prime }$ in the
above) is the most general form of uniform continuity (also of inversion of $%
F^{\prime }$), since it necessarily commutes with (category) limits. This is
one of the main key in analysis, where it comes not only as integration by
parts, but also in spectral theory: It converts the monoid $End(X)$ into a $%
\mathcal{C}^{\prime }$-function space on the spectrum $X^{\prime }$.

After these generalities worth mentioning, let us examine the case of the
forgetting functor $\Phi :Ch_{-}\rightarrow $ $Mod_{-}$ from the category $%
Ch_{-}\ $of cochain complexes $\mathcal{F}_{-}$ (as above) to the category $%
Mod_{-}$ of negatively graded modules: The functor $\Phi $ simply forgets
boundary operators absent in $Mod_{-}$, or equivalently sets them equal to $%
0 $. To attain a unified treatment of the notion of \textit{contractibility}
for complexes with various kinds of algebraic structures, we need the
following\newline

\noindent \textbf{Definition 1}: A cochain complex $\mathcal{F}_{-}$ (as
above) is \textit{contractible} if and only if it is $Mod_{-}$-free or
-cofree. We denote such by $\tilde{0}$: $\mathcal{F}=\ \tilde{0}$.\newline

The term \textit{contractible} is slightly ambiguous because in the usual
chain theory it means homotopic to $0$ (cofree), while in the space theory
it means homotopic to a point (free and cofree), and in the algebraic
cochain theory it means homotopic to scalar (free). But these various
meanings can be treated in quite a coherent manner as free or cofree object.

Before explaining our use of the definition, let us mention useful
intermediate cases, like being $Set_{-}$-free in $Mod_{-}$, or forgetting $%
\partial $ beyond some fixed degree. The former case associates to each
subchain complex $\mathcal{F}^{\prime }\subset \mathcal{F}$ a semi-direct
decomposition $\mathcal{F}\approx \mathcal{F}^{\prime }+\mathcal{F}^{\prime
\prime }$, which is actually isomorphic to a direct one if one factor is $%
\tilde{0}$ (whose proof is easy). The second case describes acylicity above
the chosen degree. \newline

\noindent \textbf{Definition 2}: Two cochain complexes $\mathcal{F}_{-},\ 
\mathcal{F}_{-}^{\prime }$ are homotopy equivalent if and only if $\mathcal{F%
}_{-}\oplus {\tilde{0}}\approx \mathcal{F}_{-}^{\prime }\oplus {\tilde{0}%
^{\prime }}$.\newline

To make more explicit the meaning of this notion, and to see its implication
at the chain map level, recall the mapping cylinder $cyl(\phi )$ of a
cochain map $\phi :\mathcal{F}_-\rightarrow \mathcal{F}^{\prime }_-$ of
degree $0$. This chain construction mimics the graph of a map, as embedding
into a simplex with vertices in either domain or range, and arrows $%
x\longrightarrow \phi (x)$: 
\begin{eqnarray}
&\qquad & \qquad \quad \ \downarrow  \nonumber \\
cyl(\phi )&:&\begin{diagram} \node{cyl(\phi )_{-k}\ \ \ =} \arrow{s}
\node{\mathcal{F}_{-k}} \arrow{s,t}{\partial_{-k}} \node{\oplus \ \ \quad
\mathcal{F}_{-k+1} \ \ \quad \oplus} \arrow{s,r}{\partial_{-k+1}}
\arrow{se,t}{\phi_{-k+1}} \arrow{sw,l}{(-1)^{k-1}} \node{\mathcal{F}'_{-k}}
\arrow{s,r}{\partial'_{-k}} \\ \node{cyl(\phi )_{-k+1}\ \ =}
\node{\mathcal{F}_{-k+1}} \node{\oplus \ \ \quad \mathcal{F}_{-k+2} \ \
\quad \oplus} \node{\mathcal{F}'_{-k+1}} \end{diagram}\ \ .  \nonumber \\
&\qquad & \qquad \quad \ \downarrow
\end{eqnarray}
\noindent The subcomplexes, $\mathcal{F}_-$ and $\mathcal{F}^{\prime }_-$,
in $cyl(\phi )$ have their quotients $cone(\phi )$ and $cone(id_{\mathcal{F}%
})=cone(\mathcal{F}_-)$, respectively: 
\begin{eqnarray}
&&\qquad \quad \downarrow  \nonumber \\
cone(\phi ) &:&\begin{diagram} \node{cone(\phi)_{-k}} \arrow{s} \node{=}
\node{\mathcal{F}_{-k+1}} \arrow{s,l}{\partial_{-k+1}}
\arrow{ese,t}{\phi_{-k+1}} \node{\oplus} \node{\mathcal{F}'_{-k}}
\arrow{s,r}{\partial'_{-k}} \\ \node{cone(\phi)_{-k+1}} \node{=}
\node{\mathcal{F}_{-k+2}} \node{\oplus} \node{\mathcal{F}'_{-k+1}}
\end{diagram}\ \,  \nonumber \\
&&\qquad \quad \downarrow
\end{eqnarray}

\begin{eqnarray}
&&\qquad \quad \quad \downarrow  \nonumber \\
cone(id_{\mathcal{F}})=cone(\mathcal{F}) &:&\begin{diagram}
\node{cone(\mathcal{F})_{-k}=} \arrow{s} \node{\mathcal{F}_{-k}}
\arrow{s,l}{\partial_{-k}} \node{\oplus} \node{\mathcal{F}_{-k+1}}
\arrow{s,r}{\partial_{-k+1}} \arrow{wsw,t}{(-1)^{k-1}} \\
\node{cone(\mathcal{F})_{-k+1}=\ \ } \node{\mathcal{F}_{-k+1}} \node{\oplus}
\node{\mathcal{F}_{-k+2}} \end{diagram}\ \ \ .  \nonumber \\
&&\qquad \quad \quad \downarrow
\end{eqnarray}

\noindent \textbf{Lemma}: The forgetting functor $\Phi :Ch_{-}\rightarrow
Mod_{-}$ has a left adjoint functor $\Phi ^{l}:Mod_{-}\rightarrow Ch_{-}$
and a right adjoint functor $\Phi ^{r}:Mod_{-}\rightarrow Ch_{-}$.\newline

\noindent Proof: From any given complex $M_{-}=\{M_{k}\}$ of negatively
graded modules (i.e., $M_{k}=0$ for $k>0$), we can construct a cochain
complex $\Phi ^{l}(M_{-})$ belonging to $Ch_{-}$ by defining $\Phi
^{l}(M)_{-k}:=M_{-k-1}\oplus M_{-k}$ together with the boundaries $\Phi
^{l}(M_{-})_{-k}=M_{-k-1}\oplus M_{-k}\stackrel{\left[ 
\begin{array}{cc}
0 & (-)^{k}I \\ 
0 & 0
\end{array}
\right] }{\longrightarrow }M_{-k}\oplus M_{-k+1}=\Phi ^{l}(M_{-})_{-k+1}$,
which terminates with $\Phi ^{l}(M_{-})_{0}=M_{-1}\oplus M_{0}$ (not with $%
M_{0}$!) at degree $0$. By chasing the commutativity in the diagram
(starting from $k=1$): 
\begin{equation}
\begin{array}{cccccc}
\Phi ^{l}(M) & :\ \ \rightarrow & M_{-k-1}\oplus M_{-k} & \stackrel{{\tiny %
\left[ 
\begin{array}{cc}
0 & (-)^{k}I \\ 
0 & 0
\end{array}
\right] }}{\rightarrow } & M_{-k}\oplus M_{-k+1} & \rightarrow \\ 
\downarrow &  & ^{\alpha _{-k}}\searrow \swarrow ^{\beta _{-k}} &  & 
^{\alpha _{-k+1}}\searrow \swarrow ^{\beta _{-k+1}} &  \\ 
\mathcal{F}^{\prime } & :\ \ \rightarrow & \mathcal{F}_{-k}^{\prime } & 
\stackrel{{\partial }_{-k}^{\prime }}{\rightarrow } & \mathcal{F}%
_{-k+1}^{\prime } & \rightarrow
\end{array}
,
\end{equation}
each chain map $\Phi ^{l}(M_{-})\rightarrow \mathcal{F}_{-}^{\prime }$ is
seen to correspond bijectively to an arbitray choice of module maps $\beta
_{-k}:M_{-k}\rightarrow \mathcal{F}_{-k}^{\prime }$ for $k>0$ through the
relations 
\begin{equation}
{\partial }_{-k}^{\prime }\beta _{-k}=(-)^{k}\alpha _{-k+1}\ \text{for }k>0 
\nonumber
\end{equation}
which determine the $\alpha _{-k}$'s. The cochain $\Phi ^{l}(M_{-})$ is
related to the preceeding cone on a cochain by means of the shift functor in 
$Mod_{-}$ : $(sM_{-})_{-k}\equiv M_{-k+1}$: With the $0$ coboundary on $%
M_{-} $, $\Phi ^{l}(sM_{-})=cone(M_{-})$. For the right adjoint, just define 
$\Phi ^{r}(M_{-})=cone(M_{-})$.%
\endproof%
%

\noindent \newline
\textbf{Remark}: Taking $\beta _{-k}=I=id$ with$\ M_{-}=\Phi (\mathcal{F})$
and $\mathcal{F}_{-}^{\prime }=\mathcal{F}_{-}$ in the above, we see that
the counit $\epsilon _{\mathcal{F}_{-}}:(\Phi ^{l}\circ \Phi )(\mathcal{F}%
_{-})\rightarrow \mathcal{F}_{-}$ is given by the projection cochain map $%
(\epsilon _{\mathcal{F}})_{-k}=((-)^{k+1}\partial _{-k-1},\ I):\mathcal{F}%
_{-k-1}\oplus \mathcal{F}_{-k}\longrightarrow \mathcal{F}_{-k}$. As for the
unit $\mathcal{F}_{-}\rightarrow (\Phi ^{r}\circ \Phi )(\mathcal{F}_{-})$,
it is the cochain inclusion $\mathcal{F}_{-k}\rightarrow \mathcal{F}%
_{-k}\oplus \mathcal{F}_{-k+1}$. \newline

The importance of the shift functor $s$ is well known in topology to express
the topological suspension operator as a quotient of the cylinder $M\times
\lbrack 0,1]$ with each basis collapsed into one point (or, both together
with one edge $m_{0}\times \lbrack 0,1]$ into the same point). The
suspension can also be an open cone, when the half open interval is used
instead. However, its crucial role here will be to express the regular
representation of a (Lie) algebra. It appears as the second component in the
above $cone(\mathcal{F}_{-})$. To stress the shift from a $0$-dimensional
object to a $1$-dimensional arrow we shall also use the famous bar notation $%
\overline{M}$ for $sM$ and $\overline{Mod}_{-}$ for the image category of
strictly negatively graded modules. Let $s,\ cone^{\prime },\
cone:Ch_{-}\rightarrow Ch_{-}$ be the canonical extension of $s,\Phi
^{l},\Phi ^{r}:Mod_{-}\rightarrow Ch_{-}$. Then the relation $\Phi ^{l} (%
\overline{M})=\Phi ^{r}(M)$ extends to $cone^{\prime}(\overline{\mathcal{F}%
_{-}})=cone(\mathcal{F}_{-})$.

\noindent \newline
\textbf{Theorem 1}: $\Phi ^{r}\Phi \approx $ $cone$ by the cochain map $\phi 
$ corresponding to the $Mod_{-}$-map $\Phi (\mathcal{F}_{-})\rightarrow \Phi
(cone^{\prime }(\mathcal{F}_{-}))$. Similarly, $\Phi ^{l}\Phi \approx
cone^{\prime }$ by the cochain map corresponding to the $Mod_{-}$-map $\Phi
(cone(\mathcal{F}_{-}))\rightarrow \Phi (\mathcal{F}_{-}).$\newline

\noindent Proof : Defining $\phi _{-k}$ for each $k$ by 
\begin{eqnarray}
\phi _{-k}:{\Phi }^{l}\Phi ({\overline{\mathcal{F}}})_{-k}= &\Phi ^{r}\Phi (%
\mathcal{F})_{-k}&={\overline{\mathcal{F}}}_{-k-1}\oplus {\overline{\mathcal{%
F}}}_{-k}=\mathcal{F}_{-k}\oplus \mathcal{F}_{-k+1}  \nonumber \\
&\stackunder{{\tiny \left[ 
\begin{array}{cc}
I & 0 \\ 
\partial _{-k} & (-)^{k}I
\end{array}
\right] }}{\longrightarrow }&\mathcal{F}_{-k}\oplus \mathcal{F}_{-k+1}=cone(%
\mathcal{F}_{-})_{-k}
\end{eqnarray}
we see that it satisfies $\phi _{-k+1}\partial _{{\Phi }^{l}\Phi ({\overline{%
{\mathcal{F}}_{-}}})}=\partial _{cone({\mathcal{F}}_{-})}\phi _{-k}$, where $%
(\partial _{cone(\mathcal{F}_{-})})_{-k}=\left[ 
\begin{array}{cc}
\partial _{-k} & (-)^{-k}I \\ 
0 & \partial _{-k+1}
\end{array}
\right] $ and $(\partial _{\Phi ^{l}\Phi ({\overline{\mathcal{F}_{-})}}%
})_{-k}:=\left[ 
\begin{array}{cc}
0 & (-)^{-k}I \\ 
0 & 0
\end{array}
\right] .$ Thus we obtain a cochain map $\phi :\Phi ^{l}\Phi ({\overline{%
\mathcal{F}_{-}}})\rightarrow cone(\mathcal{F}_{-})$ which provides the
alluded isomorphism 
\begin{equation}
\begin{diagram} \node{\Phi^l \Phi({\overline {\mathcal{F}_-}}): }
\arrow{s,l}{\phi} \node{\rightarrow} \node{\mathcal{F}_{-k-1}\oplus
\mathcal{F}_{-k}} \arrow{s,l}{\tiny {\left[ \begin{array}{cc} I & 0 \\
\partial _{-k-1} & (-)^{k+1}I \end{array} \right]}} \arrow[3]{e,t}{\tiny
{\left[ \begin{array}{cc} 0 & (-)^{k+1}I \\ 0 & 0 \end{array} \right]}}
\node{} \node{} \node{\mathcal{F}_{-k}\oplus \mathcal{F}_{-k+1}}
\arrow{s,l}{\tiny {\left[ \begin{array}{cc} I & 0 \\ \partial _{-k} &
(-)^{k}I \end{array} \right] }} \node{\rightarrow}\\
\node{cone(\mathcal{F}_-):} \node{\rightarrow}
\node{\mathcal{F}_{-k-1}\oplus \mathcal{F}_{-k}} \arrow[3]{e,b}{\tiny \left[
\begin{array}{cc} \partial_{-k-1} & (-)^{k+1}I \\ 0 & \partial _{-k}
\end{array} \right] } \node{} \node{} \node{\mathcal{F}_{-k}\oplus
\mathcal{F}_{-k+1}} \node{\rightarrow} \end{diagram}.
\end{equation}
The other isomorphism results from this one by the above relation $\Phi ^{l}(%
\overline{M_{-}})=\Phi ^{r}(M)$. 
\endproof%
%
\newline

Note that this natural isomorphism $\phi $ cannot be extended beyond the
linear structure, namely, it will not respect bilinear compositions involved
in algebraic structures.

Now we can see that $cyl(\phi )\approx cone(\mathcal{F})\oplus \mathcal{F}%
^{\prime }\approx \mathcal{F}^{\prime }\oplus \tilde{0}$ : The projection $%
cyl(\phi )\rightarrow cone(\mathcal{F}_{-})$ has a cochain section by using
the freeness of $cone^{\prime }({\overline{\mathcal{F}_{-}})}=cone(\mathcal{F%
}_{-})$ and the module map ${\overline{\mathcal{F}_{-}}}\rightarrow cyl(\phi
)$. Moreover $\phi $ factors canonically through $\mathcal{F}%
_{-}\hookrightarrow cyl(\phi )\rightarrow \mathcal{F}_{-}^{\prime }$, the
last map being a homotopy equivalence as defined before. Combined with the
other inclusion $\mathcal{F}_{-}\hookrightarrow cyl(\mathcal{F}_{-})$, it
generates homotopy of maps: Two cochain maps $\mathcal{F_{-}%
\rightrightarrows F}_{-}^{\prime }$ are homotopic iff they factor through
the two canonical inclusions $\mathcal{F}_{-}\rightrightarrows cyl(\mathcal{F%
}_{-})\stackrel{h}{\longrightarrow }\mathcal{F}_{-}^{\prime }$ by some chain
map $h$. Then it has been shown \cite{deRhKervMaum67} that the homotopy
equivalence of cochain complexes as defined previously is equivalent to the
existence of a pair of opposite chain maps, with both compositions being
homotopy equivalent to the identity.

From the above existence of free and cofree objects, we see that homotopy is
a selfdual relation. Then it applies both to negatively graded cochains $%
\mathcal{F}_{-}$ and to their positively graded dual $\mathcal{F}^{+}$. For
instance, $cone(\mathcal{F}^{+})$ becomes free, while $cone^{\prime }(%
\mathcal{F}^{+})$ becomes cofree. Both type of cochains come together in
duality (see Sect.4).

The classical graded commutative example is the de Rham complex of
differentiable $p$-forms and differentiable currents on a manifold $M$
(called by him even and odd forms) corresponding to an orientation $\omega $
of the orientable cover. More generally, we can define the space $\mathcal{F}%
_{k}\equiv \Omega ^{n-k}.\omega $ of differentiable currents w.r.t. any
fixed $0$-current $\omega $ with its support being the whole manifold. It is 
$\Omega ^{0}=C^{\infty }$-(pre-)dual to the de Rham algebra of differential
forms $\Omega \equiv \mathcal{F}^{+}$.\newline

\noindent \textbf{Theorem}: A Riemannian metric transforms any $p$-tangent
field into a differentiable current $\in \mathcal{F}_{-p}$, and provides a $%
\Omega ^{0}$-isomorphism $\mathcal{F}_{-p}\approx \mathcal{F}^{p}$.\newline

The latter Hodge duality $\mathcal{F}_{-k}\approx \mathcal{F}^{k}$ equips $%
\mathcal{F}_-$ and $\mathcal{F}^+$ with both boundary and coboundary $d =
\partial, d^{\ast }=\ast \partial \ast $. So they become mixed cochain
complexes. The invariance of the Hamiltonian $H$ by $d, d^{\ast } $ implies $%
H=dd^{\ast }+d^{\ast }d$ = Laplacian.

In the present work, however, we shall use a Taylor superexpansion of a de
Rham complex (see Sect.4). Contrary to the case of de Rham complex, it is of
infinite degree but degreewise finite dimensional. It displays linearly the
nonlinear part of the de Rham complex.

In this way we come to the point to focus upon a \textit{positively} graded
cochain complex $\mathcal{F}^{+}$ put in a duality relation with a
negatively graded one $\mathcal{F}_{-}$. In this situation the coboundary $d$
on $\mathcal{F}^{+}$ is seen to be \textit{implemented} by the boundary $%
\partial $ acting on currents in $\mathcal{F}_{-}$ in the following sense:
For $\forall f\in \mathcal{F}^{+}$ and $\forall c\in \mathcal{F}_{-}$, we
have $\partial (f\cdot c)=df\cdot c+(-1)^{f}f\cdot \partial (c)$, or
equivalently, $[\partial ,\ f\cdot _{\mathcal{F}_{-}}]c=df \cdot _{\mathcal{F%
}_{-}}c$, which means 
\begin{equation}
[\partial ,\ f\cdot _{\mathcal{F}_{-}}]=df\cdot _{\mathcal{F}_{-}}.
\end{equation}
Combined with the multiplication $f\cdot _{\mathcal{F^{+}}}$of $\ f\in 
\mathcal{F}^{+}$ on $\mathcal{F}^{+}$ (distinguished from the one $f\cdot _{%
\mathcal{F}_{-}}$on $\mathcal{F}_{-}$), this equation also implies the
derivation property of $d$ on $\mathcal{F}^{+}$, $d(fg)=(df)g+(-1)^{f}f(dg)$%
, for $f,\ g\in \mathcal{F}^{+}$. Rewriting the latter into 
\begin{align*}
[d,\ f\cdot _{\mathcal{F^{+}}}]=df\cdot _{\mathcal{F^{+}}},
\end{align*}
we see that the multiplication in $\mathcal{F}^{+}$, $(f,\ g)\mapsto
fg=f\cdot _{\mathcal{F^{+}}}g=\mu (f\otimes g),$ is a cochain map, $\mu :%
\mathcal{F}^{+}\otimes \mathcal{F}^{+}\rightarrow \mathcal{F}^{+}, $ of
degree $0,$ i.e., $d\circ \mu =\mu \circ (d$ $\otimes I+(-)^{\bullet
}I\otimes d)$, where $d$ $\otimes I+(-)^{\bullet }I\otimes d$ is the
coboundary operator of $\mathcal{F}^{+}\otimes \mathcal{F}^{+}$. Taking the
dual of this $\mu $, we find a coproduct structure $\mathcal{F}%
_{-}\rightarrow \mathcal{F}_{-}\otimes \mathcal{F}_{-}$ in the (pre-)dual $%
\mathcal{F}_{-}$. Note also that $\mathcal{F}^0$ is closed under the
composition, that $\mathcal{F}^{>0}$ is a $\mathcal{F}^{0}$-bimodule, and
that $\mathcal{F}^{0}\rightarrow \mathcal{F}^{>0}$ is a derivation.

The above map $\mu $ makes the cochain complex $\mathcal{F}^{+}$ an
associative algebra which we call an \textit{algebraic cochain complex}. It
can equally be replaced by a Lie bracket, which yields a \textit{%
Lie-algebraic cochain complexes:}\newline

\noindent \textbf{Definition 3}: A cochain complex $C$ is called an \textit{%
algebraic} (resp.,~\textit{Lie-algebraic}) \textit{cochain complex} under a
composition map $\mu :C\otimes C\rightarrow C$ if $\mu $ is a cochain map
satisfying the axioms of associative product (resp.,~Lie bracket). Together
with arrows as cochain maps commuting with the structure map $\mu $, the
totality of such complexes constitute a category, $AlgCh$ (resp.,~$LieAlgCh$%
), of (Lie-)\hfill \break algebraic objects in the category $Ch$ of cochain
complexes. \newline

\noindent \textit{NB: }They should not be confused with cochain complexes
having an algebra or Lie algebra at each degree.\newline

We shall soon discuss the homotopy theory of algebraic cochain complexes and
of Lie-algebraic ones. While the coboundary of an algebraic cochain complex
is a graded derivation (which can be viewed as a first variation by algebra
automorphisms), the homotopy uses algebra homomorphisms. Similar situation
is found for Lie-algebraic cochain complexes of graded derivations, whose
maps can be viewed as variations of algebra homomorphisms by inner actions
of algebra automorphisms.

To study the homotopy for a (Lie-)algebraic cochain complex in $\mathcal{C}%
=(Lie)AlgCh^{+}$ (or with negative integer grading), we consider the
forgetting functors $\Phi :AlgCh^{+}\rightarrow LieCh^{+}\rightarrow Ch^{+}$
and their composition, all denoted generically by the same symbol $\Phi $.
In contrast to the shift functor $s$ to be discussed later as a regular
representation, these functors do not change scalars. As in the above
definition, we adopt here a unified notation $\mu (a\otimes b)$ for both the
associative product $ab$ of $a,b$ in $AlgCh^{+}$ and the Lie bracket $[a,b]$
in $LieCh^{+}$ (which should be easily judged according to the context). 
\newline

\noindent \textbf{Definition 4}: A cochain complex in $\mathcal{C}$ is $%
\mathcal{C}$-\textit{contractible} if and only if it is $Ch^{+}$-free (or
cofree) with a cone cochain complex as its basis: $\mathcal{C}(\Phi
^{l}(cone),\ C)\approx Ch^{+}(cone,\ \Phi (C))$ or $\mathcal{C}(C,\ \Phi
^{r}(cone^{\prime }))\approx Ch^{+}(\Phi (C),\ cone^{\prime })$ with $\Phi
^{l}$ and $\Phi ^{r} $, the left and right adjoint functor of the forgetting
one $\Phi :\mathcal{C}\rightarrow Ch^{+}$, respectively. \newline

\noindent \textbf{Theorem 2}: There exist functors $Ch^{+}\stackrel{\mathcal{%
L}}{\longrightarrow }LieCh^{+}\stackrel{\mathcal{U}}{\longrightarrow }%
AlgCh^{+}$ together with their composition $\mathcal{A}\equiv $ $\mathcal{%
U\circ L}$, all of which are adjoint to the corresponding forgetting ones. 
\newline

\noindent Proof: Once one obtains them neglecting coboundaries, it suffices
to extend the coboundaries on generators as derivations. This reduces the
construction to the case $Mod^{+}\stackrel{\mathcal{L}}{\longrightarrow }%
Lie^{+}\stackrel{\mathcal{U}}{\longrightarrow }Alg^{+}$. The functor $%
\mathcal{U}$ is known as the universal envelopping algebra which generates
from a Lie algebra an associative algebra $Tens(M^{+})$ modulo the
identification of the Lie bracket with the algebraic commutator. The functor 
$\mathcal{L}$ is constructed as follows:

\noindent $\mathcal{L}(M^{+})=Tens(M^{+})$ modulo [scalars$=0$, antisymmetry
and Jacobi identity]

= $\mathcal{L}(M^{0})\otimes (K\oplus M^{1}\oplus (M^{2}\oplus \mu
(M^{1}\otimes M^{1}))\oplus$

$\ \ \oplus (M^{3}\oplus \mu (M^{1}\otimes M^{2})\oplus (\mu (\mu
(M^{1}\otimes M^{1})\otimes M^{1})\oplus \cdots )$ modulo [scalars$=0$]. 
\newline
It has a faithful representation in the underlying Lie algebra of the
algebra $Tens(M^{+})$, generated by $M^{+}$. Finally, the universal
enveloping algebra of a free Lie algebra on $M^{+}$ is the free algebra $%
\mathcal{A}(M^{+})=\oplus _{k=0}^{\infty }(\stackunder{j_{1}+\cdots +j_{r}=k%
}{\oplus }(M^{j_{1}}\otimes \cdots \otimes M^{j_{r}}))$ $\equiv Tens(M^{+})$
with the multiplication structure defined by the tensor product $\otimes $. 
\endproof%
%
\newline

These functors create free objects. Note that $\mathcal{U}$ applied to a
graded Lie algebra $M^{+}$ with bracket $0$ yields a $Mod^{+}$-free algebra $%
\mathcal{A}_{c}(M^{+})$ which is graded-commutative in Koszul sense. We note
that shift operation $s$, $(sA)^{k}=A^{k-1}$ destroys the graded product
structure of $A^{+}$ in$\ (Lie)AlgCh^{+}$ replacing it with the underlying
graded module structure: 
\[
s:(Lie)AlgCh^{+}\rightarrow \overline{Ch^{+}}. 
\]
Such a module structure can be regarded as a regular representation in view
of the usual identification of a representation of $A^{+}$ on $V$ with the
algebra $A\oplus V$ where $\mu (V\otimes V)=0$. As before, we use here the
bar notation $s(A^{+})=\overline{A^{+}}$. The elements of $A^{+}$ act on
this $\overline{A^{+}}$ as graded operators $A^{+}\times \overline{A^{+}}%
\rightarrow \overline{A^{+}}$.

Consider the diagram 
\[
\begin{array}{ccc}
(Lie)AlgCh^{+} & \stackrel{\mathcal{A},\mathcal{L}}{\leftrightarrows } & 
Ch^{+} \\ 
\downarrow \Phi _{d=0} &  & ^{\overline{d}=0}\downarrow \uparrow ^{cone} \\ 
(Lie)Alg^{+} & \stackrel{s}{\rightleftarrows } & \overline{Mod^{+}}
\end{array}
. 
\]
Applying $\mathcal{A}$ to $cone(M)=M^{+}\oplus \overline{M^{+}}=cone^{\prime
}(\overline{M})$ we get an algebraic sum (see the end of this section).%
\newline

\noindent \textbf{Corollary 3}: The functor $\Phi
_{d=0}:(Lie)AlgCh^{+}\rightarrow (Lie)Alg^{+}$ forgetting the coboundary $%
d=0 $ has a left adjoint functor $\Phi _{d=0}^{l}$ producing $(Lie)AlgCh^{+}$%
-cones.\newline

\noindent Proof: For $Alg^{+}\rightarrow AlgCh^{+}$ we define it by $\Phi
_{d=0}^{l}:=\mathcal{A}{}\circ cone^{\prime }\circ s={}\mathcal{A}\circ
cone\circ \Phi $ modulo the relations $a\otimes b=ab$ in the given graded
algebra $A$. This gives an algebraic sum $A\otimes \mathcal{A}(\overline{A})$
(ibid.), which has the required property. For $Lie^{+}\rightarrow LieCh^{+}$%
, the relation are $a\otimes b-(-)^{ab}b\otimes a=[a,b]$. Then the algebraic
sum is $Lie-cone(L)=L\otimes \mathcal{L}(\overline{L})$ by virtue of
graded-commutativity.%
\endproof%
%
\newline

We note, however, that de Rham differential cochain complexes are
characterized by $\Omega ^{0}$-freeness over the regular $(Lie)Alg$
representation of the space of vector fields on the space $\Omega _{-1}$ of
differentiable $(-1)$-currents. It means to introduce furthermore the
relation $a\otimes \overline{b}=\overline{ab}$ or $[a,\overline{b}]=%
\overline{[a,b]}$ between an algebra $A^{+}$ or $L^{+}$ and its regular
representation $\overline{A^{+}}$ or $\ \overline{L^{+}}$. This gives de
Rham cone $A^{+}\otimes ^{n}{\mathcal{A}}(\overline{A})$ and $L^{+}\otimes 
\mathcal{L}(\overline{L})$. As will be discussed in Sect.4, the Taylor
expansion homotopy model replaces $\Omega ^{0}$ by the ground scalars when $%
V=0$. In that case, a cone $cone(\overline{A^{+}})=\overline{A^{+}}\oplus (s%
\overline{A^{+}})$ will be the Weil algebra \cite{Cart50}.

Now we relate the above discussion to the bar construction. As in the charge
formula, the principle is to transform (derivative) operations into interior
mutiplications. Abstractly, this is the induction process to extend $K$%
-scalars to larger $U$-scalars: Given a $K$-algebra $U$ with rank one
representation $\epsilon :U\rightarrow K$, the induction $U\otimes M$
provides a $U$-module with the action of $U$ as an internal operation. These 
$U$-modules are the $Mod$-free ones (w.r.t. the operation-forgetting
functor). The corresponding resolution $U\otimes M\rightarrow M$ has a
kernel, coinciding with the image of $\overline{U}\otimes M$ with $\overline{%
U}=Ker\epsilon $. Starting again with $M$ replaced by the latter and so on,
gives the bar resolution $B(U,M)$ by a $Mod$-free chain augmented by $M$ in
degree -1, equal to $U\otimes (\otimes ^{n}\overline{U})\otimes M$ in degree 
$n$. It has even a canonical linear homotopy $I\sim 0$. Then the
representation $\overline{U}$ can be replaced by the full regular one, as $%
sU $. Now $B(U,U)$ can be interpreted as de Rham current complex of a
classifying space for $U$-action, and so any birepresentation of $U$ on a
vector space $S$ defines the de Rham cochain of the base space with values
in $S$ as $B(U,U)^{\ast }\otimes S\approx Hom_{U}(B(U,U),S)$. Its cohomology
is called cohomology of $U$ with coefficient $S$. The explicit form of its
coboundary is found in \cite{MacLane63}.\newline

\noindent \textbf{Definition 5}: Two algebraic cochain complexes in a
category $\mathcal{C}$ are $\mathcal{C}$-homotopy equivalent iff they become
isomorphic by adding $\mathcal{C}$-acyclic algebraic cochain complexes $%
\tilde{0}$.\newline

Here the adding operation requires explanation. It refers to a category sum,
whose definition is to be given by a coherent bijection $\mathcal{C}(X 
\stackunder{\mathcal{C}}{+}Y,Z)\approx \mathcal{C}(X,Z)\times \mathcal{C}%
(Y,Z)$. However, such a sum does not always exist; it may exist for certain
objects but not all of them. For instance, free objects $FX^{\prime }$
w.r.t.~a functor $F^{\prime }:\mathcal{C}\rightarrow \mathcal{C}^{\prime }$
to a category $\mathcal{C}^{\prime }$ with sum, have a $\mathcal{C}$ sum: 
\begin{eqnarray}
\mathcal{C}(F(X^{\prime }\stackunder{\mathcal{C}^{\prime }}{+}Y^{\prime }),\
Z) &\approx &\mathcal{C}^{\prime }(X^{\prime }\stackunder{\mathcal{C}%
^{\prime }}{+}Y^{\prime },\ F^{\prime }Z)\approx \mathcal{C}^{\prime
}(X^{\prime },F^{\prime }Z)\times \mathcal{C}^{\prime }(Y^{\prime
},F^{\prime }Z)\newline
\nonumber \\
&\approx &\mathcal{C}(FX^{\prime },Z)\times \mathcal{C}(FY^{\prime },Z).
\end{eqnarray}
Any object $X\in \mathcal{C}$ has the free presentation $FF^{\prime }X%
\stackrel{\varepsilon _{X}}{\rightarrow }X$, with relation $RX$. In
algebraic cases, this gives the sum $X\stackunder{\mathcal{C}}{+}Y\approx
FF^{\prime }X\stackunder{\mathcal{C}}{+}FF^{\prime }Y$, modulo $RX,\ RY$. Of
course $F^{\prime }(X\stackunder{\mathcal{C}}{+}Y)\approx F^{\prime }X%
\stackunder{\mathcal{C}^{\prime }}{+}F^{\prime }Y$.

In the case $\mathcal{C}=Alg^{+}\rightarrow Mod^{+}=\mathcal{C}^{\prime }$,
we have the partially defined sum $Tens(M^{+})\stackunder{Alg^{+}}{+}%
Tens(N^{+})=Tens(M^{+}\oplus N^{+})$ in $Alg^{+}$. For the subcategory $%
gcAlg^{+}$ of graded-commutative algebras, it becomes a totally defined sum $%
\wedge (M^{+})\stackunder{gcAlg^{+}}{+}\wedge (N^{+})=\wedge (M^{+}\oplus
N^{+})\approx (\wedge (M^{+}))\otimes (\wedge (N^{+}))$. In the
corresponding infinitesimal case $\mathcal{C}=Lie^{+}$ of Lie-bracket
composition, we have the partially defined sum $\mathcal{L}(M^{+})%
\stackunder{Lie^{+}}{+}\mathcal{L}(N^{+})\approx \mathcal{L}(M^{+}\oplus
N^{+})$. In the first and third cases, the partially defined sum extends
totally by resolution technique.

Now we can use the above definition to find $\mathcal{C}$-cylinder with $%
\mathcal{C}$ among these algebraic categories. Namely, $cyl(X)\approx X%
\stackunder{\mathcal{C}}{+}cone(X)$, with $\mathcal{C}$-cone, and the two
inclusions of $X$ in the first and second component. This will define $%
\mathcal{C}$-homotopy as we said, as well as $\mathcal{C}$-homotopy
equivalence. It will be used to deform a given physical supersymmetry data
(a priori only even-odd graded) to the one which is integer graded. Then the
new data carries the real homotopy type of the configuration space.

\section{Principal Algebraic Cochain Complexes}

At each step of an expansion we will encounter the appearance of some
symmetry groups, which requires us to treat a principal bundle $P$ with a
symmetry group $G$ as its structure group. The data of $P$ amounts to a $G$%
-equivariant map $\epsilon :P\rightarrow P_{G}$ to a principal $G$-bundle $%
P_{G}$ with a contractible total space, called a \textit{universal} bundle.
We start by describing the abstract algebraic structure of the de Rham
cochain algebra $\mathcal{F}^{+}\equiv \Omega ^{+}(P)$ along the line of 
\cite{GrueHalpVans3}. Dually we could use the de Rham coalgebra $\Omega
_{-}(P)$.

\begin{description}
\item[1)]  $\mathcal{F}^{+}$ is an algebraic cochain complex, either $\Bbb{Z}%
_{2}$ (even-odd)- or $\Bbb{N}$-graded,

\item[2)]  $\frak{g}$ is a (classical) Lie algebra acting on the $Alg$%
-cochain complex $(\mathcal{F}^{+},d)$ by a $0$-derivation $\theta $
commuting with $d$ defined by $\theta _{X}=\partial _{s}\exp (sX)|_{s=0}$,

\item[3)]  the vertical action of $\exp X\in G$ comes with a cochain
algebraic homotopy $dh_{X}+h_{X}d=I-\exp (sX)$ where $h_{X}$ integrates
along the path $\exp (tX),0\leq t\leq 1$. Hence $h_{X}$ commutes with $%
\theta _{X}$. The infinitesimal vertical homotopy $\iota _{X}=\partial
_{s}h_{sX}|_{s=0}$ is a ($-1$)-derivation of $\mathcal{F}^{+}$. By
differentiation of the above homotopy relation, we have the Cartan identity 
\begin{equation}
\lbrack d,\iota _{X}]=d\iota _{X}+\iota _{X}d=\theta _{X}.  \label{Cartan}
\end{equation}
It is not an algebraic cochain homotopy, but only a linear homotopy, since $%
\theta _{X}$ does not preserve the algebraic structure. These three maps $%
\iota _{X},\ \theta _{X},\ d$ are now derivations of degree $-1,\ 0,\ +1$.
In the $\Bbb{Z}_{2}$-graded case, $\pm 1$ reduces to $odd\equiv +1\equiv -1\
(\hbox{\rm mod.}2)$. Since $\iota _{X}$ is constant along the flow generated
by $X$, we have also 
\begin{equation}
(\iota _{X})^{2}=0,
\end{equation}
and since $\exp [X,Y]$ is the commutator of $\exp X$ and $\exp Y$,$\ $we
have 
\begin{equation}
\lbrack \theta _{X},\iota _{Y}]=\iota _{\lbrack X,Y]}.
\end{equation}
\ (However, any $h$ commuting with $\theta $ would give such an operation by
taking $\iota =hdh$). %
\end{description}

\noindent \textbf{Proposition}: The infinitesimal homotopy $\iota $
integrates to one for $\exp X$.\newline

\noindent Proof : The exponential $\exp (t\theta _{X})$ of the $0$%
-derivation $t\theta _{X}$ is an algebraic automorphism. Because of its
spectral decomposition into two self annihilating endomorphisms $td\iota
_{X}+t\iota _{X}d$, we have 
\[
\exp (t\theta _{X})=\exp (td\iota _{X})\exp (t\iota _{X}d)=\exp (td\iota
_{X})+\exp (t\iota _{X}d)-I. 
\]
Taking the variation between $t=0,\ 1$, this gives linearly 
\begin{equation}
I-\exp (\theta _{X})=dh_{X}+h_{X}d  \nonumber
\end{equation}
with $h$ an algebraic cochain homotopy in the categorical sense defined
before. 
\endproof%
%
\newline

Now a $G$-connection $A$ on $P$ provides another homotopy equivalent cochain
subcomplex (one of the Brown models) compatible with $\theta $ and $\iota $,
consisting of the $G$-invariant (scalar) $p$-forms. It is the model $\Omega
(P)_{\iota =0}\otimes \wedge \frak{g}^{1}\hookrightarrow \mathcal{F}$ in the
algebraic form. Neither factor of $\Omega (P)_{\iota =0}$, $\wedge \frak{g}%
^{1}$ is a subcomplex, and $d$ has a component $A:\frak{g}^{\ast
}\rightarrow \mathcal{F}^{1}$. The base algebraic subcomplex is $\Omega
(Z)\approx \mathcal{F}_{\iota =0,\theta =0}$. The commutation with $\iota $
of the associated algebra inclusion $\wedge \frak{g}^{1}\hookrightarrow 
\mathcal{F}$ yields $\iota _{X}(A(X^{\ast }))=\iota _{X}(X^{\ast })=\langle
X,\ X^{\ast }\rangle $ at lowest degrees $1$ and $0$. This decomposition $%
\Omega (P)_{\iota =0}\otimes \wedge \frak{g}^{1}$ is the degree $1$ analog
to the degree $0$ induction $\Omega ^{0}(Z)\subset \Omega ^{0}(P)\approx
\Omega ^{0}(Z)\otimes Sym\frak{{g_{0}}}$, which gives the Taylor expansion
of the commutative algebra $C(G;\Omega ^{0}(Z))$ for the trivial action
equipped with a pointwise product. The Poisson algebra consisting of Noether
charges is now augmented in degree $0$ by $\frak{g}$. Its quantization is
the Heisenberg-Yang-Lie superalgebra $L_{-}\oplus (\hslash \Bbb{C}\oplus 
\frak{g})_{0}\oplus L^{+}$.

A fixed connection $A$ provides a classifying object in the sense that $%
\mathcal{F}^{+}=\Omega ^{+}(P)$ has a classifying algebraic cochain map $%
\epsilon _{A}:W_{\frak{g}}\rightarrow \mathcal{F}^{+}$ from the Weil algebra 
$W_{\frak{g}}$. To explain this fundamental notion is worth a short
digression back to the categorical adjunction: A category $Cat$ has an
adjunction with the point category (on one object with its identity) iff one
of its objects $I$ (the free one) satisfies the condition that each object $%
X $ correponds bijectively to a morphism $\epsilon _{X}:I\rightarrow X$ such
that $f\epsilon _{X}=\epsilon _{Y}$ for $\forall f:X\rightarrow Y$. Then, $%
Cat$ is said to be contractible, and it looks like a cone. Any category $%
\mathcal{C}$ gives rise to contractible ones in the following way: For an
object $A\in Ob(\mathcal{C})$ fixed, the category $\mathcal{\hat{C}}$ [$=$ $%
(A\downarrow \mathcal{C})$: the category of objects under $A$] is
contractible when it is defined by $Ob(\mathcal{\hat{C})}:=\{h\in Cat(A,\
X); $ $X\in Ob(\mathcal{C})\mathcal{\}}$ and by $\mathcal{\hat{C}}%
(h,k)=\{u_{\ast }\equiv u\circ (-)\ ;\ u\in Cat(X,\ Y)$ s.t.$\ k=u\circ h\}$
as the set of morphisms from $h\in Cat(A,\ X)\ $to $k\in Cat(A,Y)$. The
reason is because each object $h\in Ob(\mathcal{\hat{C})}$ corresponds
bijectively to a morphism $h_{\ast }$ in $\mathcal{\hat{C}}$ from $I_{A}$%
(=the identity of $A$) to $h:$ $h=h_{\ast }(I_{A})=h\circ I_{A}$. This is
just the essence of the so-called Yoneda's song. But it is nothing but a
cone construction identifying objects with arrows, which naturally involves
a dimension shift by $1$.

Returning to our data, the inclusion of $\frak{g}$-algebra $\wedge \frak{g}%
^{1}\hookrightarrow \mathcal{F}^{+}$ given by the above connection
corresponds to an $AlgCh$-map $\epsilon _{A}:Alg-cone(\wedge \frak{g}%
^{1})\rightarrow \mathcal{F}^{+}$. This is the classifying map saying that
the category of principal algebraic cochain complexes is adjoint to the
point category. But in the category of graded-commutative algebras, this
cone $Alg-cone(\wedge \frak{g}^{1})$ is nothing but the $\Bbb{N}$-graded
Weil algebra $W_{\frak{g}}^{+}=Sym(\frak{g}^{2})\otimes \wedge \frak{g}^{1}$%
, which has been seen with the notation $\overline{L}=\frak{g}^{1},\ s%
\overline{L}=\frak{g}^{2}$ and with the assignment of a $0$ superbracket to $%
\frak{g}^{1}\oplus \frak{g}^{2}$. This approach has the advantage of showing
the a priori freeness, in particular the $Cat$-contractibility. This is
usually expressed only as acyclicity, but in our context, the length
operator is homotopic to $0$, because the extension of the base cochain
linear homotopy $dk+kd=I$ as odd-derivations $\overline{d},\overline{k}$
gives a derivation $[\overline{d},\overline{k}]$ equal to the identity on
generators. But the extension $\overline{I}$ is length $1$. Now the
classifying map $\epsilon _{A}:W_{\frak{g}}^{+}\rightarrow \mathcal{F}^{+}$
expresses $A$ as push out, $A=\epsilon _{A}(W)$, of the unique (non-flat)
Weil connection $W:\frak{g}^{\ast }\rightarrow W_{\frak{g}}^{1}=\frak{g}%
^{\ast }$.

Our new step is to introduce a principal Lie chain basis for $\Omega (P):$ 
\begin{equation}
L(P):\ \cdots \rightarrow L_{-2}(P)\rightarrow L_{-1}(P)\rightarrow
L_{0}(P)\rightarrow P_{+1}\rightarrow 0,
\end{equation}
where $P_{+1}=Z_{+1}\oplus \frak{g}_{+1}$. It projects onto $L(Z)$ with
kernel $L(G)$ augmented by $\frak{g}_{1}$. The homology exact sequence of $%
0\rightarrow L(G)\rightarrow L(P)\rightarrow L(Z)\rightarrow 0$ will be
(Hurewicz-)isomorphic to the homotopy exact sequence of the fibration. The
bracket $[\frak{g}_{1},\ L_{-p}]\subset L_{-p+1}$ generates the derivation $%
\iota $ on $\Omega (P)$. The map $L_{0}(P)\rightarrow Z_{+1}\oplus \frak{g}%
_{+1}$ consists of two components taking values in $Z_{+1}$ and in $\frak{g}%
_{+1}$, each of which can be identified, respectively, with the first order 
\textit{solderin}g 1-form and the first order \textit{connection} 1-form.
Similarly, the composition $L_{-1}(P)\rightarrow L_{0}(P)\rightarrow
Z_{+1}\oplus \frak{g}_{+1}$ has the corresponding \textit{torsion} and 
\textit{curvature} as its two components, both of which are $0$. In the next
section, we construct these Lie chain bases.

\section{Taylor Superexpansion of de Rham Complex}

In this section we show how graded Taylor coefficients $\partial
_{i_{0}}\cdots \partial _{i_{q}}f_{\mu _{1}\cdots \mu _{p}}\ dz^{\mu
_{1}}\wedge $ $\cdots \wedge dz^{\mu _{p}}(\in J^{q}\Omega ^{p})$ of $p$%
-forms $\sum\limits_{\mu _{1}\cdots \mu _{p}}f_{\mu _{1}\cdots \mu _{p}}\
dz^{\mu _{1}}\wedge $ $\cdots \wedge dz^{\mu _{p}}\in \Omega ^{p}$ are
related to the homotopy group $\pi _{q+1}(Z)$ of $\alpha :\Bbb{R}^{q+1}\cup
\{\infty \}\rightarrow Z$ which sends $\infty $ to a fixed base point of a
manifold $Z$ under consideration.

A first observation is the duality between the space of forms and the
differential algebra acting on the former by taking the Taylor coefficients
of $\alpha ^{\ast }f$ at the origin $\in \Bbb{R}^{q+1}$. We shall use the
Weil bicomplex expanding the de Rham complex over an open cover \cite{Weil52}
to produce a new bicomplex model exhibiting the Hurewicz map $\pi
_{k+1}(Z)\rightarrow H_{-k-1}(Z)$. Here $Z$ is assumed to be a real analytic
manifold, the typical examples of which are given by Lie groups and their
homogeneous spaces. On $Z$ let us choose a cover $\mathcal{R}=\{U_{i}\}$ by
open domains of analytic charts, such that each non empty intersection $%
U_{i_{0}\cdots i_{q}}\equiv U_{i_{0}}\cap \cdots \cap U_{i_{q}}$ has a base
point named as $i_{0}\cdots i_{q}$ not belonging to a higher order
intersection. Choose also a partition of unity $\{\varphi _{i}\},\ \sum
\varphi _{i}=1$ subordinate to the cover $\{U_{i}\}$: Namely, each $\varphi
_{i}$ is a real analytic function $(>0)$ (typically of such a Gaussian shape
as constant$\times \exp (\frac{-\varepsilon ^{2}}{\varepsilon ^{2}-r^{2}})$,
see, e.g., \cite{Schw66} for details) in the interior of $U_{i}(\simeq \Bbb{R%
}^{n})$ which extends smoothly (but not analytically, of course) as $0$ to
the outside of $U_{i}$. Now let $\Omega _{\mathcal{R}}^{0}$ be the algebra
of functions of the form $f=\stackunder{i_{0}\cdots i_{q}}{\sum }\varphi
_{i_{0}}\cdots \varphi _{i_{q}}a_{i_{0}\cdots i_{q}}$, where $a_{i_{0}\cdots
i_{q}}$ are analytic functions on $U_{i_{0}\cdots i_{q}}\subset Z$ and each
term in the sum is differentiably extended by $0$ on $Z$. By the choice of
the base points, $a_{i_{0}\cdots i_{q}}$ is determined by the Taylor
expansion of $f$ at $i_{0}\cdots i_{q}$. Hence we have a linear
decomposition $\Omega _{\mathcal{R}}^{0}\approx \stackunder{i_{0}\cdots i_{q}%
}{\oplus }\Omega ^{0}(U_{i_{0}\cdots i_{q}})$, where each summand denotes
the space of analytic functions. On the manifold $Z$, $\Omega _{\mathcal{R}%
}^{0}$ generates the corresponding $\Omega _{\mathcal{R}}^{0}$-modules of
tangent vector fields $\mathcal{X}_{\mathcal{R}}=Der(\Omega _{\mathcal{R}%
}^{0})$, of differentiable ($-1$)-currents $\Omega _{-1}^{\mathcal{R}%
}\approx \mathcal{X}_{\mathcal{R}}$, and of cotangent fields $\Omega _{%
\mathcal{R}}^{1}=\Gamma _{\mathcal{R}}(T^{\ast }(Z))$ as the $\Omega _{%
\mathcal{R}}^{0}$-dual of $\Omega _{-1}^{\mathcal{R}}$, and algebraic
cochain complexes, $\Omega _{-}^{\mathcal{R}}=\stackunder{p\in \Bbb{N}}{%
\oplus }\stackunder{i_{0}\cdots i_{q}}{\oplus }\Omega _{-p}(U_{i_{0}\cdots
i_{q}}),\ \Omega _{\mathcal{R}}^{+}=\stackunder{p\in \Bbb{N}}{\oplus }%
\stackunder{i_{0}\cdots i_{q}}{\oplus }\Omega ^{p}(U_{i_{0}\cdots i_{q}})$,
of \v{C}ech type in the indices $i_{0}\cdots i_{q}$ and of de Rham type in $%
p $, in a similar way to the de Rham one, as explained in Sect.1: $\Omega
_{-p}(U_{i_{0}..i_{q}})\equiv \stackrel{p}{\stackunder{\Omega _{\mathcal{R}}%
}{\wedge }}\Omega _{-1}(U_{i_{0}\cdots i_{q}}),\ \Omega ^{p}(U_{i_{0}\cdots
i_{q}})\equiv \stackrel{p}{\stackunder{\Omega _{\mathcal{R}}}{\wedge }}%
\Omega ^{1}(U_{i_{0}\cdots i_{q}})$.\newline

\noindent \textbf{Theorem 4}: If $\mathcal{R}$ is convex, the complex $%
\Omega _{\mathcal{R}}^{+}$ is weakly homotopy equivalent (i.e., as linear
cochains) to the de Rham one.\newline

\noindent Proof: Here we take $\mathcal{R}$ to be a de Rham convex cover 
\cite{deRh50, Weil52} and use the Poincar\'{e} lemma on a convex open set,
and the (bi)complex technique, in particular, the fact that a short exact
sequence $0\rightarrow C^{\prime }\rightarrow C\rightarrow C^{\prime \prime
}\rightarrow 0$ is split when $C^{\prime }\sim 0$ or $C^{\prime \prime }\sim
0$ \cite{deRhKervMaum67}. Let $\partial /\partial t$ be the radial field on
the pointed convex set $U_{i_{0}\cdots i_{q}}$. The canonical cone operator
defined by $h(f^{(k)})(X_{(k-1)})=\int_{0}^{1}f^{(k)}(t\ X_{(k-1)}\wedge
\partial /\partial t)dt$ for $(k-1)$-vector fields $X_{(k-1)}$ and $k$-forms 
$f^{(k)}$ satisfies $hd=I-dh$. It respects analyticity, hence is a homotopy $%
I\sim 0$ on $\Omega ^{+}(U_{i_{0}\cdots i_{q}})$. 
\endproof%
%
\newline

The new analytic model $\Omega _{\mathcal{R}}^{+}$ for the de Rham complex
contains much deeper information in view of the familiar formula describing
the action of the coboundary $d=d^{\prime }+d^{\prime \prime }$ on a form $%
\varphi \in \Omega _{\mathcal{R}}^{+}$ in terms of vector fields $X_{i}$: 
\begin{eqnarray}
(d^{\prime }\varphi )(X_{1},\cdots ,X_{p+1}) &=&\stackunder{i}{\sum }%
(-)^{i+1}[X_{i},\varphi (X_{1},\cdots ,\stackrel{\vee }{X_{i}},\cdots
,X_{p+1})],  \nonumber \\
(d^{\prime \prime }\varphi )(X_{1},\cdots ,X_{p+1}) &=&\stackunder{i<j}{\sum 
}(-)^{i+j}\varphi ([X_{i},X_{j}],X_{1},\cdots ,\stackrel{\vee }{X_{i}}%
,\cdots ,\stackrel{\vee }{X_{j}},\cdots X_{p+1}),  \nonumber
\end{eqnarray}
where $[X,\ f]=\partial _{X}f=Xf$ for functions $f$.

While the space $\Omega _{-p}(Z)=\stackrel{p}{\stackunder{\Omega ^{0}}{%
\wedge }}\Omega _{-1}(Z)$ of currents, in $\Omega ^{0}$-duality to $p$-forms 
$\Omega ^{p}(Z)$, is the $p$-th power of $\Omega _{-1}(Z)$ w.r.t. $\Omega
^{0}$-linear exterior product $\stackunder{\Omega ^{0}}{\wedge }$ over $%
\Omega ^{0}$, we now consider $\wedge ^{p}\Omega _{-1}(Z)$ taking the $\Bbb{R%
}$-linear exterior product $\wedge \equiv \stackunder{\Bbb{R}}{\wedge }$
over $\Bbb{R}$. In contrast to the usual current $c_{X_{1}\wedge \cdots
\wedge X_{p}}\in \Omega _{-p}(Z)$ with point-like localization, an element $%
c_{X_{1}}\wedge \cdots \wedge c_{X_{p}}\in \wedge ^{p}\Omega _{-1}(Z)$ is
not localized at a single point in the base manifold $Z$, but it involves $p$
tangent vectors $X_{i}\in T_{z_{i}}(Z)$ at $p$ different points $z_{i}\in Z$%
. In view of the canonical isomorphism $T_{(z_{1},\cdots
,z_{p})}(Z^{p})\approx \stackrel{p}{\stackunder{i=1}{\oplus }}T_{z_{i}}(Z)$
and of the canonical embedding $\wedge ^{p}(\stackrel{p}{\stackunder{i=1}{%
\oplus }}T_{z_{i}}(Z))\subset \wedge ^{p}T_{(z_{1},\cdots ,z_{p})}(Z^{p})$,
we have an isomorphism $\wedge ^{p}\Omega _{-1}(Z)\approx \Omega
_{-p}(Z^{p}-Diag_{p}(Z))^{sym}$ based upon $c_{X_{1}}\wedge \cdots \wedge
c_{X_{p}}=(c_{X_{1}}\wedge \cdots \wedge c_{X_{p}})(z_{1},\cdots ,z_{p})=%
\stackunder{\sigma =(\sigma _{1},\cdots ,\sigma _{p})\in \frak{S}_{p}}{\sum }%
X_{\sigma _{1}}(z_{1})\wedge \cdots \wedge X_{\sigma _{p}}(z_{p})$, since
the last expression vanishes for $(z_{1},z_{2},\cdots )\in Diag_{p}(Z)\equiv
\{(x_{1},x_{2},\cdots )\in Z^{p};x_{i}=x_{j}\ \mathrm{for\ some\ }%
i\not=j\}\subset Z^{p}=\stackrel{p}{\overbrace{Z\times \cdots \times Z}}$.
In comparison, we have $(c_{X_{1}}\wedge \cdots \wedge c_{X_{p}})(z,\cdots
,z)=X_{1}(z)\wedge \cdots \wedge X_{p}(z)=c_{X_{1}\wedge \cdots \wedge
X_{p}}(z)$. A map $\alpha :\Bbb{R}^{p}\rightarrow Z$ gives $\wedge
^{p}\Omega ^{1}(Z)\rightarrow \wedge ^{p}\Omega ^{1}(\Bbb{R})\approx \Omega
^{p}(\Bbb{R}^{p}-Diag_{p}(\Bbb{R}))^{sym}\stackrel{\cong }{\rightarrow }\Bbb{%
R}$ and hence an element $c_{\alpha }\in \wedge ^{p}\Omega _{-1}$. It can be
visualized as a smoothing homotopy of $\alpha :\Bbb{R}^{p}\rightarrow
Diag_{p}(Z)$ pushing it away from $Diag_{p}(Z)$ (i.e. $0$ on it; in general, 
$\alpha $ itself cannot be pushed away from $Diag_{p}(Z)$). This element $%
c_{\alpha }$ behaves as a local current on $Z^{p}$ under a deformation of $%
\alpha $ supported around a point. This is the relevance of $\wedge
^{p}\Omega _{-1}(Z)$ to homotopy, even with its bracket $[\alpha ,\beta ]$
viewed in $Z\vee Z\subset (Z\times Z-Diag_{2}(Z))\cup \{\mathrm{base\ point}%
\}$, while $\Omega _{-p}(Z)$ is relevant to the homology. Likewise, the jets
on $\Omega _{-1}(Z)$ can be defined similarly to the ordinary currents with
a Lie derivative $\partial _{Y}$ and with $J^{r}=\partial _{Y_{1}}\cdots
\partial _{Y_{r}}$ as 
\begin{eqnarray*}
(\partial _{Y}c_{X})(\varphi ) &\equiv &-c_{X}(\partial _{Y}\varphi
)=-\langle \iota _{X}\partial _{Y}(\varphi )\rangle _{\omega }, \\
J^{r}c_{X} &=&(-)^{r}c_{X}J^{r},
\end{eqnarray*}
but their meaning on $\wedge ^{p}\Omega _{-1}(Z)$ is spread as symmetric
currents over $Z^{p}-Diag_{p}(Z)$, not localized at a point on $Z$ as
ordinary currents $c_{X_{1}\wedge \cdots \wedge X_{p}}$, the latter of which
vanish for $p>n$. Keeping this in mind, we write an $r$-jets in $%
J^{r}(\wedge ^{p}\Omega _{-1}(Z))$ as $\partial _{Y_{1}}\cdots \partial
_{Y_{r}}c_{X_{1}}\wedge \cdots \wedge c_{X_{p}}$.

By the above formula $d=d^{\prime }+d^{\prime \prime }$ (which is equivalent
on $\Omega ^{1}(Z)$ to $[\partial _{X},\ \iota _{Y}]=\iota _{[X,\ Y]}$ and $%
\partial _{X}=d\iota _{X}+\iota _{X}d$), we get a boundary $\partial
:L_{-1}\rightarrow L_{0}$ as the $\Bbb{R}$-dual of $d:\Omega
^{1}(Z)\rightarrow \Omega ^{2}(Z)$ by 
\begin{equation}
\partial (c_{X}\wedge c_{Y})=-\partial _{X}c_{Y}+\partial
_{Y}c_{X}+c_{[X,Y]},\newline
\end{equation}
where $L_{0}\equiv \Omega _{-1}(Z)$ and $L_{-1}\equiv \wedge ^{2}\Omega
_{-1}(Z)$. The last term $\delta :L_{-1}\ \backepsilon \ c_{X}\wedge
c_{Y}\longmapsto \lbrack c_{X},\ c_{Y}]\equiv c_{[X,Y]}\in L_{0}$, $\Bbb{R}$%
-dual to $d^{\prime \prime }$, defines a Lie bracket $[L_{0},L_{0}]\subset
L_{0}$.

Since the derivative $\partial _{X}$ taking $1$-jets is linear only in $\Bbb{%
R}$ but not in $\Omega ^{0}$, the canonical quotient map $\wedge ^{p}\Omega
_{-1}(Z)\rightarrow \Omega _{-p}(Z)=\stackunder{\Omega ^{0}}{\stackrel{p}{%
\wedge }}\Omega _{-1}(Z)$ is not compatible with the $1$-jet operation, so
the right-hand side of the above formula is not defined over the quotient
space $\Omega _{-p}(Z)$. The above $\partial $ respects the $r$-jet
filtration: $\partial (J^{r}\Omega _{-1})\subset J^{r+1}\Omega _{-1}$ and $%
[J^{r}\Omega _{-1},J^{s}\Omega _{-1}]\subset J^{r+s}\Omega _{-1}$. The role
of this jet filtration is just to make the spaces $L_{0},L_{-1}$
jet-degreewise finite-dimensional by taking Taylor expansions at a chosen
discrete set of base points in $\mathcal{R}$. This ensures a reflexive
duality between $\Omega ^{1}$ and $\Omega _{-1}$.

The above structure generates a Lie algebraic chain complex $L\equiv
\{L_{-p}\}_{p\in \Bbb{N}_{0}}$ with 
\begin{equation}
L_{-p}\equiv \wedge ^{p+1}\Omega _{-1},\newline
\end{equation}
which is augmented by the evaluation $L_{0}\rightarrow L_{1}=J^{0}\Omega
_{-1}$ $\approx $ tangent space at a base point. On this complex a mapping $%
d_{L}:L_{-p}\longrightarrow L_{-p+1}$ is defined by 
\begin{equation}
d_{L}(c_{X_{1}}\wedge \cdots \wedge c_{X_{p+1}})\equiv \stackunder{i}{\sum }%
(-)^{i+1}\partial _{X_{i}}c_{X_{1}}\wedge \cdots \wedge \stackrel{\vee }{%
c_{X_{i}}}\wedge \cdots \wedge c_{X_{p+1}},
\end{equation}
which is $\Bbb{R}$-dual to the first component $d^{\prime }$ in $d$ in the
limit of coincident points: 
\begin{eqnarray*}
&&(d_{L}(c_{X_{1}}\wedge \cdots \wedge c_{X_{p+1}}))(\varphi )=\stackunder{i%
}{\sum }(-)^{i+1}\langle \partial _{X_{i}}\varphi (X_{1}\wedge \cdots \wedge 
\stackrel{\vee }{X_{i}}\wedge \cdots \wedge X_{p+1})\rangle _{\omega } \\
&=&\langle (d^{\prime }\varphi )(X_{1}\wedge \cdots \wedge X_{p+1})\rangle
_{\omega }=\langle c_{X_{1}\wedge \cdots \wedge X_{p+1}}(d^{\prime }\varphi
)\rangle _{\omega }.
\end{eqnarray*}
Similarly, a Lie superbracket $[L_{-p},L_{-q}]\subset L_{-p-q}$ is defined
as a natural extension of the above $\delta :L_{-1}\rightarrow L_{0}$ by 
\begin{eqnarray}
&&[c_{X_{1}}\wedge \cdots \wedge c_{X_{p+1}},c_{Y_{1}}\wedge \cdots \wedge
c_{Y_{q+1}}]  \nonumber \\
&\equiv &\stackunder{i,j}{\sum }(-)^{i+j+p+1}c_{[X_{i},Y_{j}]}\wedge
c_{X_{1}}\wedge \cdots \wedge \stackrel{\vee }{c_{X_{i}}}\wedge \cdots
\wedge c_{X_{p+1}}\wedge  \nonumber \\
&&\qquad \wedge c_{Y_{1}}\wedge \cdots \wedge \stackrel{\vee }{c_{Y_{j}}}%
\wedge \cdots \wedge c_{Y_{q+1}},
\end{eqnarray}
whose coincidence point limit is $\Bbb{R}$-dual to $d^{\prime \prime }$. By
derivation rule we also have $[J^{r}L_{-p},J^{s}L_{-q}]\subset
J^{r+s}L_{-p-q}$ as 
\begin{eqnarray}
&&[J^{r}c_{X_{1}}\wedge \cdots \wedge c_{X_{p+1}},J^{s}c_{Y_{1}}\wedge
\cdots \wedge c_{Y_{q+1}}]  \nonumber \\
&\subset &J^{r+s}\stackunder{i,j}{\sum }(-)^{i+j+p+1}c_{[X_{i},Y_{j}]}\wedge
c_{X_{1}}\wedge \cdots \wedge \stackrel{\vee }{c_{X_{i}}}\wedge \cdots
\wedge c_{X_{p+1}}\wedge  \nonumber \\
&&\qquad \wedge c_{Y_{1}}\wedge \cdots \wedge \stackrel{\vee }{c_{Y_{j}}}%
\wedge \cdots \wedge c_{Y_{q+1}}.
\end{eqnarray}

Finally, to recapture $d^{\prime \prime }$ from $L$, we take the
construction $\mathcal{A}_{c}(L)$ seen in Sect.2, which gives a free
presentation of the original cochain algebra $\Omega $. This uses the
reflexivity of the duality mentionned above, so we have to use $L_{-p}$ over
the cover $\mathcal{R}$ in such a form as a double chain complex with the
total chain complex 
\begin{equation}
L_{k}\equiv \stackunder{p+q=k}{\oplus }L_{-p}(U_{i_{o}\cdots i_{q}})
\end{equation}
filtered by finite-dimensional chain complexes 
\begin{equation}
L_{k,l}\equiv \stackunder{p+q=k}{\oplus }\ \stackunder{q+r=l}{\oplus }%
J^{r}(\wedge ^{p}\Omega _{-1}(U_{i_{0}\cdots i_{q}})),  \label{Taylor}
\end{equation}
where the jets taken at the base point $i_{0}\cdots i_{q}$ constitute a
finite-dimensional chain subcomplex. Because of the constraint $p-r=k-l=%
\mathrm{const}$ (following from $p+q=k$ and $q+r=l$ in Eq.(\ref{Taylor})),
the gradation of $L_{k}$ in $p$ with $L_{-p}(Z)\approx \oplus
L_{-p}(U_{i_{o}})/L_{-p}(U_{i_{o}i_{1}})$ can be reinterpreted as a
filtration in the jet degree $r$ arising from the Taylor superexpansion of $%
\Omega _{-}(Z)$ (in anticommuting variables, each of finite dimensions).

Now $\Omega _{\mathcal{R}}$ has a resolution of the form $\mathcal{F}_{%
\mathcal{R}}=\mathcal{A}_{c}(\overline{L(Z)^{+}})$ seen in Sect.2: 
\begin{eqnarray*}
\mathcal{F}_{\mathcal{R}} &:&\mathcal{F}^{0}=Sym(Z)\otimes K\longrightarrow 
\mathcal{F}^{1}=Sym(Z)\otimes L^{0} \\
\stackrel{d}{\longrightarrow }\mathcal{F}^{2} &=&Sym(Z)\otimes (L^{1}\oplus
(L^{0}\wedge L^{0})) \\
\stackrel{d}{\longrightarrow }\mathcal{F}^{3} &=&Sym(Z)\otimes (L^{2}\oplus
(L^{0}\wedge L^{1})\oplus (\wedge ^{3}L^{0}))\stackrel{d}{\longrightarrow }%
\cdots .
\end{eqnarray*}
where $L^{p}\equiv L_{-p}^{\ast }$. One can interpret $L^{p}(U_{i_{0}\cdots
i_{q}})$ as coordinates of a supermanifold $Z^{L}$ with a body manifold $Z$
(i.e., the $0$ degree part). Then $\mathcal{F}_{\mathcal{R}}$ is the de Rham
supercomplex. This means that the new model does not lose the geometry \cite
{DeWi92}.

The former geometrico-analytic construction of $L$ based on the locality of $%
\Omega $ realizes a direct categorical construction in the $1$-connected
case, but it has more profound essence applicable to such general situations
as non-simply connected cases or $\Bbb{Z}_{2}$-graded ones. So let us
consider a degreewise finite-dimensional coalgebraic cochain complex of the
form $\mathcal{F}_{-}=\mathcal{A}_{c}(\overline{L_{-}(Z)})$ with a Lie
superalgebraic cochain complex $L_{-}(Z)$ augmented by a module $Z$,
together with its dual version, a free algebraic cochain complex over a Lie
super-coalgebra $\mathcal{L}^{+}=\mathcal{A}_{c}(\overline{L^{+}})$ equipped
with a cobracket $L^{+}\rightarrow L^{+}\oplus (L^{+}\otimes L^{+})$. In the
case of reflexive duality (e.g., when the dimensions are degreewise finite),
the correspondence between $L$ and such ${\mathcal{F}}^{+}$ is obviously a $%
Cat$ isomorphism. Then $d$ on $\mathcal{A}_{c}(\overline{L_{-}})$ has a
Noether charge in $P=L_{-}\otimes L^{+}$ by linearity of the bracket $[\cdot
\ ,\ \cdot ]$.

Now we want to show that any $1$-connected ${\mathcal{F}}^{+}$ has such a
basis $L$ up to weak homotopy equivalence. The functors ${\mathcal{A}}_{c},%
\mathcal{L}$ met in Sect.2 allow one to define ${}\mathcal{L}%
s^{-1}:coAlg_{c}Ch_{-}\rightarrow LieAlgCh_{-}$ giving the free Lie
superalgebra cochain complex on the shifted cochains, and $co{\mathcal{A}}%
_{cs}:LieAlgCh_{-}\rightarrow coAlg_{c}Ch_{-}$ where $co\mathcal{A}_{c}=%
\mathcal{A}_{c}^{\ast }$ giving the free coalgebra on the shifted cochain. 
\newline

\noindent \textbf{Theorem 5}: In the case of degreewise finite dimensions,
these functors define an adjunction 
\begin{equation}
\phi :LieAlgCh_{-}(\mathcal{L}s^{-1}(A_{-}),L_{-})\approx
coAlg_{c}Ch_{-}(A_{-},co\mathcal{A}_{cs}(L_{-}))
\end{equation}
for $0$-connected $A_{-}$ (reduced to scalar in degree $0$). \newline

\noindent Proof: This is due to the following series of adjunctions: 
\begin{eqnarray}
&&LieAlgCh_{-}(\mathcal{L}s^{-1}(A_{-}),\ L_{-})\approx
Ch_{-}(s^{-1}(A_{-}),\Phi (L_{-}))\approx Ch^{+}(sL^{+},\Phi (A^{+})) 
\nonumber \\
&\approx &Alg_{c}Ch^{+}(\mathcal{A}_{cs}(L^{+}),A^{+})\approx
coAlg_{c}Ch_{-}(A_{-},co\mathcal{A}_{cs}(L_{-})).
\end{eqnarray}
\endproof%
%
\newline

\noindent Since adjoint functors carry a free object to a free one, we also
have \newline

\noindent \textbf{Corollary 6}: Both functors ${\mathcal{A}}_{c}$ and $%
\mathcal{L}$ respect $\mathcal{C}$-cones with $\mathcal{C}$ being $%
Alg_{c}Ch^{+}$ or $coAlg_{c}Ch_{-}$.\newline

Next, we have \newline

\noindent \textbf{Theorem 7}: The unit $\eta $ and counit $\varepsilon $ of
the adjunction $\phi $ are weak homotopy equivalence (i.e., homotopy
equivalence in $Ch_{-}$). \newline

\noindent Proof : The adjunction $\Phi $ of both categories $LieAlgCh$ and $%
(co)AlgCh$ with $Ch$ seen in Sect.2 provides $Ch$-free objects, which are
respected by the adjunction $\Phi ,\mathcal{A},\mathcal{L}$. So one first
checks that $\varepsilon $ is a homology isomorphism on $Ch$-free objects,
then use $Ch$-free resolutions \cite{Tanre83}. For $Ch$-free objects, $%
H(\Phi ^{\ast }(C))$ is determined by a spectral sequence in $H(C).$ So both 
$H(\Phi ^{\ast }(C))$ and $H(L\mathcal{A}_{c}\Phi ^{\ast }(C))$ are
determined by a spectral sequence in $H(\mathcal{A}_{c}\Phi ^{\ast }(C))$.
As $\varepsilon $ comes from the identity on the latter, it induces an
isomorphism on homology. Finally, a chain map inducing a homology
isomorphism is a $Ch$-homotopy equivalence, because its cone chain is
acyclic, hence $\thicksim 0$ \cite{deRhKervMaum67}.%
\endproof%
%
\newline

Now the Lie superalgebra of homology $H_{-k}(L_{-}(Z))$ is related to that
of homotopy $\pi _{k+1}(Z)$ by the Hurewicz map $\pi _{k+1}(Z)\rightarrow
H_{-k}(L_{-}(Z))$, which is just the analytic continuation of the constant
function $1$ along a map $S^{k+1}\rightarrow Z$. In fact, by decent in $%
\Omega ^{0}(U_{i_{0}\cdots i_{q}})$ in the bicomplex any element $\in
H_{-k}(L_{-}(Z))$ represents a constant analytic continuation along a
singular closed $(k+1)$-manifold. The discussion on the Hurewicz map leads
to the case of $G$-principal bundle $P\rightarrow Z$. The equivariant
construction of $L(P)$ is straightforward from the preceeding one. In the
case of a $G$-principal space ($Z$ = point), the invariant de Rham complex
is built on $\frak{g}_{0}$. But the minimal $L$ (i.e. with $0$ coboundary)
is the odd graded module $L(G)=Prim(G)\approx \pi (G)$ (where $Prim$ denotes
the primitive space in the Hopf homology graded algebra $H(G)$).

For a general $P$, we have an exact sequence of Lie superalgebraic cochain
complexes $0\rightarrow L(G)\rightarrow L(P)\rightarrow L(Z)\rightarrow 0$,
where $L(Z)=L(P)_{\theta =0,\iota =0}$. The best case is given by a
homogeneous space $Z=G/H$. Because the Hurewicz map sends the homotopy exact
sequence for principal fibrations to the homology exact sequence for $L$
above, it is an isomorphism on $P$ iff it is so on $Z$.\newline
\newline
\noindent \textbf{Definition 6}: We call the Lie superalgebra $%
H_{-k}(L_{-}(Z))$ the real homotopy $\pi _{k+1}(Z)$ of $Z$ if $Z$ is simple
(i.e., the canonical action of $\pi _{1}(Z)$ on all $\pi _{k}(Z)$ is
trivial). Otherwise it gives the homotopy with this action trivialized. 
\newline

To justify this definition, we briefly indicate in the case of a $1$%
-connected space $Z$ how $H_{-k}(L_{-})\approx \pi _{k}(Z)$ is verified,
together with an algorithm to compute explicitly $\pi _{k}$ from the
(co)homology data of $\Omega (Z)$. The basic principle is as follows: The
(co)homology of a Lie group is algebraically free with the homotopy as its
basis, which is displayed in various odd degrees. Next, this extends to
principal bundles by means of the Koszul-Brown homotopy model for the de
Rham (co)chain complex of the total space of a principal fibration with a
Lie group as its structure group. Finally, the Postnikov sequence of any $1$%
-connected space $Z$ factorizes $Z$ through functional principal fibrations
with disjoint homotopy. [Incidentally it is interesting to note the
parallelism between the construction of this sequence and a problem in
mathematical physics appearing in extending the Doplicher-Roberts
superselection theory \cite{DoplRob90} to the situations with spontaneous
symmetry breakdown (SSB): Starting from an algebra acted upon by a compact
group $H$ of unbroken symmetry together with some data concerning a
homogeneous space $G/H$ with an \textit{unknown} $G$, one must find a
solution to an unknown larger algebra acted upon by this unknown compact
group $G$ satisfying the criterion of SSB.] In the topological context, this
tower describes the homotopy type of the space. Algebraically, the sequence
of principal fibrations is a sequence of extensions $\mathcal{A\otimes A}%
_{c}(L^{k})$ with differential $d+d_{L}$, where $d_{L}(L)\subset \mathcal{A}$%
, on which $B\cdots B(\pi _{k+1})$ acts . This is the so-called Hirsch
extensions \cite{GrifMorg81}. Then the total $L$ has $0$ boundary and they
are Taylor expansion (of degree $k\geq 1$) of a ``Galois superextension''.
(Galois extension would occur at $k=0$). They provide a minimal model with $%
\partial =0$, whose basis exhibits the homotopy.

Any $1$-connected cochain algebra is a sequence of Hirsch extensions, with
an explicit algorithm from the knowledge of cohomology. In terms of
Lie-algebraic cochain complexes Hirsch extensions correspond to Quillen
extensions {\cite{Tanre83}}.

\section{Non-Simply Connected Case}

Let $Z^{\prime }\rightarrow Z$ be a covering map (with discrete fiber), with
classifying map $Z\rightarrow B(W)$ where $W$ is the Weyl group of $\pi
_{1}^{\prime }\equiv \pi _{1}(Z^{\prime })\subset \pi _{1}(Z)$ (defined as
the normalizer of $\pi _{1}^{\prime }\mathop{\rm mod}\nolimits\pi
_{1}^{\prime }$). To simplify the discussion let us assume that $\pi
_{1}^{\prime }$ is normal, i.e., $W=\pi _{1}/\pi _{1}^{\prime }$.

The (convex) cover $\mathcal{R\bigskip }=\{U_{i}\}$ of $Z$ lifts to that of $%
Z^{\prime }$, $\mathcal{R}^{\prime }=\mathcal{R}\times W$. The bicomplex for 
$\mathcal{R}^{\prime }$ is $\Omega _{\mathcal{R}^{\prime }}\approx \Omega _{%
\mathcal{R}}\times W,$ except for the \v{C}ech coboundary $\oplus
i_{0}\rightarrow \oplus i_{0}i_{1}$ having a $W^{\prime }$-component, which
we denote by a ``translation'' twist $\Omega _{\mathcal{R}}\stackunder{t}{%
\times }W$. Similarly $L(Z^{\prime })\simeq L(Z)\stackunder{t}{\times }W$.
But we need to solve this equation in $L(Z)$. For that purpose, the
homotopically exact sequence $Z^{\prime }\rightarrow Z\rightarrow B(W)$
gives $\Omega _{\mathcal{R}}\sim \Omega _{\mathcal{R}^{\prime }}\stackunder{t%
}{\otimes }\Omega (B(W))$ and the homotopically exact sequence $W\rightarrow
P_{W}\rightarrow B(W)$ for the unversal $W$-principal bundle $P_{W}$ ($\sim $
point) gives $0\sim W\stackunder{t}{\otimes }\Omega (B(W))\sim cone(W)$. So
we obtain $L(Z)\simeq L(Z^{\prime })\stackunder{t}{\times }\overline{W}$,
where the bar indicates degree $0$ (shift of $W$ in degree $1$).

The same argument with the universal covering $\tilde{Z}\rightarrow
Z^{\prime }$ instead of $Z^{\prime }\rightarrow Z$ gives $L(Z)\sim L(\tilde{Z%
})\stackunder{t}{\times }L_{0}(\pi _{1},\pi _{1}^{\prime })$, where $%
L_{0}(\pi _{1},\pi _{1}^{\prime })=W\times \pi _{1}^{\prime }$. When both $W$
and $\pi _{1}^{\prime }$ are abelian, we get a linear expression $L(Z)\sim L(%
\tilde{Z})\stackunder{t}{\oplus }L_{0}(\pi _{1},\pi _{1}^{\prime })$, with $%
L_{0}(\pi _{1},\pi _{1}^{\prime })=\overline{\Bbb{R}W}\oplus \overline{\Bbb{R%
}\pi _{1}^{\prime }}$. We note that finite order elements in $W$ and $\pi
_{1}^{\prime }$ are then eliminated. The central series of $\pi _{1}$ gives
a canonical expansion $L(Z)\sim L(\tilde{Z})\stackunder{t}{\oplus }L_{0}(\pi
_{1}(Z))$, where $L_{0}(\pi )=\stackunder{k}{\oplus }\Bbb{R(}\pi
_{1}^{k+1}/\pi _{1}^{k})$ which is easily shown to be a canonical graded Lie
algebra.

As $\pi _{1}(Z)\approx H_{0}(L(Z))$, we have $H_{0}(L(\tilde{Z}))=0$, and
hence, $H_{0}(L(Z))\approx L_{0}(\pi _{1}(Z))$. By applying this to the
equivariant case considered in Sect.3, we get the interpretation of $%
L_{0}(P)\rightarrow P_{1}\approx Z_{1}\oplus \frak{g}_{1}$ as the first
order soldering and connection $1$-forms: They depend only on $\pi _{1}(P)$
(related to $\pi _{1}(Z)$ by the exact sequence $\pi _{1}(G)\rightarrow \pi
_{1}(P)\rightarrow \pi _{1}(Z)\rightarrow 0$, where $\pi _{1}(G)$ is
abelian).

\section{Superinduction and FP Ghosts}

Let $R=$ $\mathcal{A}_{c}(\overline{L(P)})$ be a Taylor-de Rham complex for
a $G$-principal bundle $P\rightarrow Z$ (Sect.3): The action of the Lie
algebra $\frak{g}$ on $L(P)$ can be incorporated by augmenting $L(P)$ with $%
\frak{g}_{0}=\frak{g}$ at degree $0$, which produces a new Lie algebraic
chain complex $L(P)^{\theta }$ with bracket $[\frak{g},\ L_{-k}]\subset
L_{-k}$, $[\frak{g}_{0},\ \frak{g}_{0}]\subset \frak{g}_{0}$ and
augmentation $L(P)^{\theta }$ generates the algebraic cochain complex $%
\mathcal{U(}L(P)^{\theta })\approx \frak{U}(\frak{g}_{0})\otimes \mathcal{A}%
_{c}(\overline{L(P)})$ (partially algebraic isomorphism). If $S$ is a linear
representation of $\frak{g}$, the usual induction from the fiber $G$ to $P$
gives a (Fock) space $\mathcal{F}=(R\stackunder{\frak{U}(G)}{\otimes }%
S)_{\iota =0}=(R\otimes S)_{\theta =0,\ \iota =0}$ on $Z$. Actually if $R$
is the usual de Rham complex $\Omega (P)$, $\mathcal{F}$ is the de Rham
complex $\Omega (Z)\otimes S$ of $S$-valued $p$-forms on $Z$. If $S$ is a
superrepresentation of $\frak{g}$, namely, it is also a $\wedge \frak{g}_{1}$%
-module, or rather a $W_{\frak{g}}$-module, the the superinduction is $R%
\stackunder{W_{\frak{g}}}{\otimes }S$.

But the (non-commutative) Lie algebraic chain complex $L(P)^{\theta }$ is
weakly homotopy equivalent to another such complex $L(P)+\frak{g}_{0}$ with
two successive augmentations \cite{deRhKervMaum67}, $\cdots \rightarrow
L_{-1}(P)\rightarrow L_{0}(P)\rightarrow P_{1}(=Z_{1}\oplus \frak{g}%
_{1})\rightarrow \frak{g}_{2}$. The bracket $[L_{-p},\ \frak{g}_{1}]\subset
L_{-p+1}$ generates $\iota $ in $\Omega (P)$ while $[\frak{g}_{2},\ \cdot
]=0 $. The operator $\theta =[\frak{g}_{0},\ \cdot ]$ is recovered from the
identity $\theta =d\iota +\iota d=[\iota ,d]$ where $d=\partial ^{\ast }+%
\stackunder{k}{\sum }[L_{k},\ \cdot ]^{\ast }$. The algebraic cochain
complex $\mathcal{A}_{c}(\overline{L(P)+\frak{g}_{2}})$ is isomorphic to $%
\mathcal{A}_{c}(\overline{L(P)})\otimes (\wedge \frak{g}_{1})$. For
instance, for the universal principal bundle where $L(=\frak{g}_{1}=\frak{g}%
_{2})$ is generated by FP ghost, we have $\mathcal{A}_{c}(\overline{L(P)}%
)=W_{\frak{g}}$, $\mathcal{A}_{c}(\overline{L(P)+\frak{g}_{2}})\sim \wedge 
\frak{g}_{1}$, and so the (co)homology is $H^{\ast }(\frak{g)}$ \cite{Ojim96}%
. Actually, for any principal $G$-bundle $P$ over $Z$, we get by
superinduction $\Omega (Z)\sim (\mathcal{A}_{c}(\overline{L(P)+\frak{g}_{2}})%
\stackunder{\wedge \frak{g}_{1}}{\otimes }W_{\frak{g}})_{\theta =0,\ \iota
=0}$. This means that FP ghosts give the homotopy type of the base.

Physically such $p$-forms (super-expansion) are quite natural as
Radon-Nikodym derivatives of a vector $x\in X$ w.r.t. another (e.g.,
space-time) variable $y\in Y$: The existence of such correlation derivative $%
\partial _{y}x\in \Omega (Z)\otimes S$ means that $X$ is fibered over $Y$
locally trivially with fibre $Z$. The continuity w.r.t. the parameter space $%
Y$ means that there is a decoupling $X\approx P_{Y}\stackunder{G}{\times }Z$
with a $G$-principal bundle $P_{Y}$ over $Y$, where $G$ acts on $Z$, and a
(nonlinear) $G$-connection on $P_{Y}$. The bundle $\Omega (Z_{y})\otimes S$
may not be flat. But the cohomology bundle $H(Z_{y})\otimes S$ is flat over $%
Y$. In the noncommutative (quantized) case, the difference is that instead
of fields (infinitesimal arrows) on $Z$ we have arrows and connection of
arrows. Then $\Omega (Z)$ becomes a noncommutative de Rham complex \cite
{Conn94}. This will be developed in subsequent work.

\section{Hodge Star and Spectral Decomposition}

Since our Lie-algebraic cochain complexes $L$ are degreewise
finite-dimensional, we can safely take over to the present context the
theory of Lie bialgebras \cite{CharPres94}, the details of which are skipped
here. In particular, there is a cobracket $\delta _{\_}:L_{\_}\rightarrow
L_{\_}\otimes L_{\_}$ for which $\partial :L_{-p}\rightarrow L_{-p+1}$ is
also a derivation, and $\delta ^{+}\equiv \delta _{\_}^{\ast }:L^{+}\otimes
L^{+}\rightarrow L^{+}$ defines a Lie algebraic cochain complex. A
self-duality $L\approx L^{\ast }$ gives a Hodge star $\partial ^{\dagger
}:L^{p}\rightarrow L^{p-1}$ which is a derivation for $\delta ^{+}$. On $%
\mathcal{F}^{+}=\mathcal{A}_{c}(\overline{L^{+}})$ equipped with the
derivation $d$ generated by $\partial ^{\ast }+\delta ^{+}$, and the
associated self-duality $\mathcal{A}_{c}(\overline{L^{+}})\approx (\mathcal{A%
}_{c}(\overline{L_{-}}))^{\ast }$, we have the Hodge star $d^{\dagger }$
generated by $\partial ^{\dagger }+\delta _{-}^{\dagger }$. As a sum of
degree $-1$and degree $+1$operators, this is an odd graded. Let us consider
the restriction $H_{L}:L^{+}\rightarrow \mathcal{F}^{+}$ of the Hamiltonian $%
H=dd^{\dagger }+d^{\dagger }d$ ($\in $ Poisson algebra) on $\mathcal{F}^{+}$%
. By the derivation property of $\partial $, the mixed compositions in $%
(\partial ^{\ast }+\delta ^{+})(\partial ^{\dagger }+\delta _{-}^{\dagger
})+(\partial ^{\dagger }+\delta _{-}^{\dagger })(\partial ^{\ast }+\delta
^{+})$ are shown to cancel out: $\delta ^{+}\partial ^{\dagger }+\partial
^{\dagger }\delta ^{+}=0=\delta _{-}^{\dagger }\partial ^{\ast }+\partial
^{\ast }\delta _{-}^{\dagger }$. By the Poisson relations 
\begin{equation}
\delta ^{+}\delta _{-}^{^{\dagger }}-(-)^{pq}\delta _{-}^{^{\dagger }}\delta
^{+}=\delta _{pq}
\end{equation}
we have $H_{L}=H^{\prime }+$ $N$, where $H^{\prime }=\partial ^{\ast
}\partial ^{\dagger }+\partial ^{\dagger }\partial ^{\ast }$ is the
Hamiltonian on $L$ and $N$ is the degree-counting operator. The classical
harmonic oscillator model shows that $\delta _{-}^{^{\dagger }}$ increases
the degree by $1$ if $p$ is odd, by $2$ if $p$ is even, i.e., it unwinds the
degree as $\Bbb{Z\rightarrow }$ $\Bbb{Z}_{2}$. So $[L^{p},\ L^{q}]$\ $%
\subset L^{p+q}$. In the principal case where a Lie algebra $\frak{g}$ acts
on $L$ and commutes with $H$, we have also the action of $\frak{g=g}_{0}$ on
each $L_{-q}$.

\bigskip Finally, we apply our theory to the problem of a $\Bbb{Z}_{2}$%
-graded supersymmetry $\Omega ^{(0)}\stackrel{d}{\stackunder{d}{%
\rightleftarrows }}\Omega ^{(1)}$ mentioned in \cite{Froe98}, as to which
kind of topological information can be obtained from such a supersymmetry.
There the vanishing of torsion was required, but here with the help of
homotopy model $L$ we show that such a restriction can be replaced by the
above spectral decomposition as follows. The first step is to associate a $%
\Bbb{Z}_{2}$-graded Lie-algebraic cochain complex $L_{(0)}\stackrel{\partial 
}{\stackunder{\partial }{\rightleftarrows }}L_{(1)}$: By our local expansion
theory on cover $\mathcal{R}$, it will be obtained under the condition of
locality for $\Omega $. This means that it can be displayed on $\mathcal{R}%
=\{U_{i}\}$ with local derivatives $\partial _{i}$. This is quite a natural
a priori condition. The second step is to have a spectral decomposition
w.r.t. some Hilbert structure on $\Omega $ coming from one on $L$. It means $%
H_{L}=H^{\prime }$ $+N$ as above, which provides $L$ with a $\Bbb{N}$-graded
Lie algebraic cochain complex structure. So we have reached the \newline

\noindent \textbf{Theorem 8}: A $\Bbb{Z}_{2}$-graded supersymmetry on a
local anticommutative algebra $\Omega $ with Hilbert-norm completion $%
\overline{\Omega }$ determines a homotopy type $\pi _{k}$, $k$: integer $%
\geq 0$.

\bigskip

\section*{Acknowledgements}

A joint project involving this work has started in 1998 with a support from
the Fonds Herbette of Universit\'{e} de Lausanne, and has continued in
1999-2000 with supports from the Swiss 3$^{em}$ cycle de math\'{e}matiques,
the Fonds Landry of Universit\'{e} de Lausanne and the exchange program
sponsered by Japan Society for the Promotion of Sciences (JSPS)(S-99187).
Both the authors would like to express their sincere thanks to these
supports. The first named author (S.M.) has greatly benefited from the
hospitality of RIMS, Kyoto University, during his stay. The second named
author (I.O.) is very grateful to IMA, Universit\'{e} de Lausanne, for the
hospitality extended to him during his two stays in 1998 and 1999. He has
also been partially supported by JSPS Grants-in-Aid (No.11640113). \vfill 
\eject


\begin{thebibliography}{99}
\bibitem{Nel70}  E.~Nelson: \emph{Topics in Dynamics, I: Flows},
Mathematical Notes, Princeton University Press, 1970.

\bibitem{deRhKervMaum67}  G.~de Rham, M.~A.~Kervaire and S.~Maumary: \emph{%
Torsion et {T}ype {S}imple d'{H}omotopie}, Lecture Notes in Mathematics
No.48, Springer, 1967.

\bibitem{Cart50}  H.~Cartan: Notion d'alg\`ebre diff\'erentielle;
application aux groupes de Lie et aux vari\'et\'es o\`u op\`re un groupe de
Lie, pp.15--27 in \emph{Colloque Topologique}, 1950; La transgression dans
un groupe de Lie at dans un espace fibr\'e principal, pp.57--71 in \emph{%
Colloque Topologique}, 1950.

\bibitem{MacLane63}  S.~Mac Lane: \emph{Homology}, Springer,
Berlin-Heidelberg-New York, 1963 (1st ed.), 1975 (3rd corrected printing).

\bibitem{GrueHalpVans3}  W.~Greub, S.~Halperin, and R.~Vanstone: \emph{%
Connections, Curvature, and Cohomology}, vol.~3, Academic Press, New
York-San Francisco-London, 1976.

\bibitem{Weil52}  A.~Weil: Sur les th\'eor\`emes de de Rham, \emph{%
Commentarii Mathematici Helvetici} \textbf{26} (1952) 119--145 .

\bibitem{Schw66}  L.~Schwartz: \emph{Th\'eorie des Distributions}, Hermann
et Cie, 1966.

\bibitem{deRh50}  G.~de Rham: Complexes a automorphismes et hom\'eomorphie
diff\'erentiable, \emph{Annales de l'Institut Fourier} \textbf{2} (1950)
51--67.

\bibitem{DeWi92}  B.~DeWitt: \emph{Supermanifolds}, Cambridge Univ.~Press,
Cambridge, 1992.

\bibitem{Tanre83}  D.~Tanr\'e: \emph{Homotopie Rationelle: Mod\`eles de
Chen, Quillen, Sullivan}, Lecture Notes in Mathematics No.~1025, Springer,
1983.

\bibitem{DoplRob90}  S.~Doplicher and J.~E.~Roberts: Why there is a field
algebra with a compact gauge group describing the superselection structure
in particle physics, \emph{Comm.~Math.~Phys.} \textbf{131} (1990) 51--107.

\bibitem{GrifMorg81}  P.~A.~Griffiths and J.~W.~Morgan: \emph{Rational
Homotopy Theory and Differential Forms}, vol.~16 of \emph{Progress in
Mathematics}, Birkh\"auser, Boston-Basel-Stuttgart, 1981.

\bibitem{Ojim96}  I.~Ojima: BRS Symmetry as a Fundamental Principle in
Quantum Gauge Theories, pp.47--62 in M.~Abe, N.~Nakanishi and I.~Ojima,
eds., \emph{{``\textit{BRS Symmetry}''}, Proc. of the International
Symposium on BRS Symmetry}, Universal Academy Press, 1996; N.~Nakanishi and
I.~Ojima: \emph{Covariant Operator Formalism of Gauge Theories and Quantum
Gravity}, World Scientific Publ.~Co., 1990.

\bibitem{Conn94}  A.~Connes: \emph{Non Commutative Geometry}, Academic
Press, 1994(Expanded English translation).

\bibitem{CharPres94}  V.~Chari and A.~Pressley: \emph{A Guide to Quantum
Groups}, Cambridge University Press, 1994.

\bibitem{Froe98}  J.~Fr\"ohlich, O.~Grandjean and A.~Recknagel:
Supersymmetric quantum theory, non-commutative geometry, and gravitation,
pp.221--385 in A.~Connes, K.~Gawedzki and J.~Zinn-Justin, eds., \emph{Les
Houches, Session LXIV, 1995: Quantum Symmetries}, Elsevier Science B.V.,
1998.
\end{thebibliography}
\end{document}